\documentclass[11pt,a4paper,iop,apj,twocolappendix]{emulateapj}
\usepackage{epsfig}
\begin{document}
\title{The Surface Brightness--Color Relations Based on Eclipsing Binary Stars: Toward Precision Better than 1\% in Angular Diameter Predictions.}
\author{Dariusz Graczyk\altaffilmark{1,2,3}, Piotr Konorski\altaffilmark{4}, Grzegorz Pietrzy{\'n}ski\altaffilmark{3,2},Wolfgang Gieren\altaffilmark{2,1}, Jesper Storm\altaffilmark{5}, \\ Nicolas Nardetto\altaffilmark{6}, 
Alexandre Gallenne\altaffilmark{7}, Pierre F. L. Maxted\altaffilmark{8}, Pierre Kervella\altaffilmark{9,10} \\ and Zbigniew Ko{\l}aczkowski\altaffilmark{11}} 
\affil{$^1$Millennium Institute of Astrophysics (MAS), Chile}
\affil{$^2$Universidad de Concepci{\'o}n, Departamento de Astronomia, Casilla 160-C, Concepci{\'o}n, Chile; darek@astro-udec.cl}
\affil{$^3$Centrum Astronomiczne im. Miko{\l}aja Kopernika (CAMK), PAN, Bartycka 18, 00-716 Warsaw, Poland; darek@ncac.torun.pl}
\affil{$^4$Obserwatorium Astronomiczne, Uniwersytet Warszawski, Al.~Ujazdowskie 4, 00-478, Warsaw, Poland}
\affil{$^5$Leibniz-Institut f\"{u}r Astrophysik Potsdam, An der Sternwarte 16, 14482 Potsdam, Germany}
\affil{$^6$Universit\'e C$\hat{\rm o}$te d'Azur, Observatoire de la C$\hat{\rm o}$te d'Azur, CNRS, Laboratoire Lagrange, UMR7293, Nice, France}
\affil{$^7$European Southern Observatory, Alonso de C{\'o}rdova 3107, Casilla 19001, Santiago 19, Chile}
\affil{$^8$Astrophysics Group, Keele University, Staffordshire, ST5 5BG, UK}
\affil{$^9$Unidad Mixta Internacional Franco-Chilena de Astronom{\'i}a (CNRS UMI 3386), Departamento de Astronom{\'i}a,
Universidad de Chile, Camino El Observatorio 1515, Las Condes, Santiago, Chile}
\affil{$^{10}$LESIA (UMR 8109), Observatoire de Paris, PSL Research University, CNRS, UPMC, Univ. Paris-Diderot, 5 place Jules Janssen,
92195 Meudon, France}
\affil{$^{11}$Instytut Astronomiczny, Uniwersytet Wroc{\l}awski, Kopernika 11, 51-622 Wroc{\l}aw,  Poland}

\begin{abstract}
In this study we investigate the calibration of surface brightness--color (SBC) relations 
based solely on eclipsing binary stars. We selected a sample of 35 detached eclipsing binaries with
trigonometric parallaxes from {\it Gaia} DR1 or {\it Hipparcos}, whose absolute dimensions are known
with an accuracy better than 3\% and that lie within 0.3 kpc from the Sun.  For the purpose of this study, we
used mostly homogeneous optical and near-infrared photometry based on the
Tycho-2 and 2MASS catalogs. We derived geometric angular diameters for
all stars in our sample with a precision better than 10\%, and for 11 of
them with a precision better than 2\%. The precision of individual angular diameters of the
eclipsing binary components is currently limited by the precision of the geometric
distances ($\sim$5\% on average). However, by using a subsample of systems
with the best agreement between their geometric and photometric distances,
we derived the precise SBC relations based only on
eclipsing binary stars. These relations have precisions that are comparable to the
best available SBC relations based on interferometric angular diameters,
and they are fully consistent with them. With very precise {\it Gaia} 
parallaxes becoming available in the near future, angular diameters with a precision better than 1\% will be abundant.
At that point, the main uncertainty in the total error budget of the SBC relations
will come from transformations between different photometric systems,
disentangling of component magnitudes, and for hot OB stars, the main uncertainty will come from the
interstellar extinction determination. We argue that all these issues can be overcome 
with modern high-quality data and conclude that
a precision better than 1\% is entirely feasible. 

\end{abstract} 

\section{Introduction}
The surface brightness--color (SBC) relations
play a fundamental role in predicting angular diameters of stars and
serve as an almost perfect tool for deriving precise distances to eclipsing
binary stars. They have also been extremely useful in Baade--Wesselink
techniques to determine the distances to classical Cepheid stars
\citep[e.g.][]{gie95,fou97,sto11}. The SBC relations are commonly
calibrated based on direct stellar angular diameters measured by means of
ground-based interferometry \citep[e.g.][]{ker04,diB05,cha14,boy14}. The
precision of the SBC relations is gradually improving thanks to the ever-growing
number of stars with interferometric angular diameters and to improvements
in dealing with the limb darkening.

Eclipsing binaries with known trigonometric parallaxes can also be used
to derive the SBC relation. This idea was first
formulated and used by \cite{lac77}: the combination of a geometric
distance and stellar radius immediately provides an angular diameter of
a component of an eclipsing binary. This can later be used to derive a
dependence of the radiative flux scale on color, expressed in terms of the
surface brightness parameter or the effective temperature. Deriving
angular diameters of eclipsing binary stars is significantly more
complex than determining an angular diameter of a single star with
interferometry. However, using eclipsing binaries usually has an important
advantage: good control on the limb-darkening uncertainties, at least when the light 
curves are of sufficient quality \citep[e.g.][]{pop84}. Early attempts
were constrained to the color $V\!-\!R$ \citep{lac77,bar78,pop80}
and were based on only on three eclipsing binary systems with secure
trigonometric parallaxes.

When Hipparcos parallaxes became available, \cite{pop98} analyzed 14 well-detached 
eclipsing binaries with the most accurate parallaxes and absolute
dimensions to compare radiative flux scales defined by interferometry
and eclipsing binary systems. However, this analysis was made only for
$(B\!-\!V)$ color and included eclipsing binaries with a significant
amount of chromospheric activity. Nonetheless, \cite{pop98} concluded
that the SBC relation based on non-active eclipsing
binaries seemed to be complementary to that based on interferometric
angular diameters. \cite{kru99} developed the idea of using
a large number of eclipsing binaries with geometric distances from
Hipparcos to precisely calibrate the SBC relations. They compiled an
extensive list of promising eclipsing binaries in the solar neighborhood
(up to 200 pc). Soon after, \cite{sem01} derived the SBC relation from
a sample of 13 eclipsing binary stars with {\it Hipparcos} parallaxes and
Str\"omgren photometry. The calibration was made for the ($b-y$) color and
compared with the relation by \cite{pop98}, which was mostly based on interferometric
and lunar occultation angular diameter measurements. The samples agreed well, but 
the derived SBC relation had very large scatter.

The usefulness of eclipsing binaries for distance measurements was
investigated by \cite{jer01} by comparison of corrected trigonometric
parallaxes and photometric distances, with the conclusion that EBs are
excellent standard candles. \cite{sma02} used 15 eclipsing binary stars
with Hipparcos parallaxes to derive a fundamental temperature scale for
A-type stars, and \cite{bil08} presented a brief analysis of using
eclipsing binary stars to calibrate the absolute magnitudes of stars as
a function of some intrinsic colors. The most recent application of
eclipsing binaries to derive the SBC relations known to us is the work by
\cite{bon06}, where the SBC relation was calibrated against $(V\!-\!K)$
color, but these authors used photometric distances to derive angular
diameters (see our Sec.~\ref{sec:ang}). Recently, \cite{sta16a,sta16b}
used more than 100 eclipsing binaries to investigate possible systematics
in recent Gaia DR1 parallaxes \citep{gaia16} and concluded that a likely
systematic shift of $-$0.25 mas is presented in Gaia parallaxes. The
shift is consistent with the systematic global error of 0.3 mas in the
DR1 that was announced by the Gaia team.

The list of eclipsing binary systems reported by \cite{kru99} is the basis for
our programme of investigating eclipsing binaries and deriving the SBC
relations. The first paper from our program was devoted to the IO~Aqr system
\citep{gra15} and showed that unrecognized triples may bias the derivation
of the SBC relations. Although the maximum-light contribution of the
third component of IO Aqr is low and relatively well determined, the SBC calibrations 
would have substantial problems to reach a precision of
about 1\% for this system. In a following paper \citep{gal16} we derived a very precise
orbital parallax to TZ~For that allowed us to perform a preliminary check
of the precision of existing SBC relations. Our parallax measurement
to TZ~For is in perfect agreement with the photometric distance and
the Gaia DR1 parallax. The work on TZ~For is a part of our larger effort to
determine very precise dynamical parallaxes to a number of long-period
eclipsing binaries.

Here we present in detail the method of deriving the SBC relations
based on eclipsing binary stars, and for the first time, we publish the
precise relations that are based solely on eclipsing binaries. Sect.~\ref{sample}
characterizes a sample of systems and describes the selection criteria and data
we used. In Sect.~\ref{method} we present the method outline of our
analysis. Sect.~\ref{results} contains results, and these are discussed in Sect.~\ref{discu}. 
The last section is devoted to final remarks.

\section{The Sample}\label{sample}
For the purpose of our work, we use
a volume-limited ($d<300$ pc) sample of detached eclipsing binaries with
published high-quality light curve and radial velocity solutions. The
sample is supposed to contain {\it standard} eclipsing binary systems
for the purpose of accurate distance determination/validation and
surface brightness calibration. We made an extensive search for suitable
systems in the literature using the SIMBAD database \citep{wen00} and
NASA ADS. Useful guidance in this measure is provided by the compilations
done by \cite{kru99}, \cite{bil08}, \cite{tor10} and more recently
by \cite{eke14}, \cite{sou15} and \cite{sta16a}. The final sample contains 34 systems and additionally AL Ari, 
a system for which our new analysis is as-yet unpublished \citep{kon17}. Our
intention is that the sample would serve as a reference catalogue for
very precise determinations of the photometric distances, the angular
diameters and the surface brightness. We put very
strict conditions for including an eclipsing binary in our sample. As
a part of a selection procedure we did an extensive consistency check
of published physical parameters for every candidate system and in some
cases recalculated fundamental parameters to make them more concordant
with the observables. Table~\ref{tab:sample} presents the basic information about
selected eclipsing binaries. The criteria are described in details below.
  
\subsection{Proximity effects} 
{\it No proximity effects larger than 0.03 mag.} Although semi-detached
or even contact configuration eclipsing binaries were suggested as good
distance indicators \citep[e.g.][]{wyi02,wil10}, our experience shows
that their physical parameters are usually much more model dependent and
thus less robust than those coming from analysis of detached eclipsing
binaries. In fact only well-detached systems offer very simple geometry
where both stars can be treated as almost perfect spheres. This simplifies
the analysis, as magnitudes and colors of the system are virtually constant
outside eclipses.

\subsection{Intrinsic variability} 
{\it No intrinsic variability amplitude larger than 0.04 mag.} Larger
variability (spots, pulsations) over a given threshold may lead to some
bias in the estimation of true photometric indices on a level of $>$2\%, so
we removed all systems with an active or pulsating component from our
sample (e.g. RS CVn stars,  chromospheric activity). The only system
retained is EF~Aqr showing some spot activity on a secondary but only
small changes in the combined out-of-eclipse light \citep{vos12}.

\subsection{Absolute dimensions}
{\it Precision better than 3\%.} For the purpose of surface brightness
calibration a knowledge of the physical radii is fundamental because
combined with a distance it gives the angular diameters. We chose known
systems with the most precise absolute dimensions. An average precision of
the radii determination in the sample is $\sigma_R/R=1.2\%$. This sample is
useful for utilization of the present Gaia parallaxes. For the future Gaia
releases expected to have precision better than 1\% for all stars in our
sample \citep{deB15} some of the systems will need to be reanalyzed in
order to achieve precision better than $2\%$ of radii determination or
eventually will have to be removed from the sample i.e. V1229~Tau, FM~Leo,
FL~Lyr, MY~Cyg, VZ~Cep, V821~Cas.  

\subsection{Geometric distance}
{\it Precision better than 10\% within 300 pc horizon}. We used trigonometric
parallaxes from the recent Gaia Data Release 1 \citep{gaia16}, augmented
with Hipparcos parallaxes \citep{vLe07} for some bright and nearby
systems. Even so, there are just a few eclipsing binaries in our sample
with high precision trigonometric parallaxes 
(fractional uncertainty $\sigma_\pi/\pi < 2\%$). In the case of one system, TZ~For, we utilized the
orbital parallax determined by \cite{gal16} which is by a factor of 5
more precise than the Gaia DR1 parallax.  

\subsection{Temperature}
{\it Effective temperatures known to within 5\%.} 
We use them to build
precise models of the systems and to calculate infrared light ratios. In
this work we utilized also temperatures to derive photometric distances
(by the flux scaling) as proxies of the true geometric distances. In
general temperatures are important for determining auxiliary parameters
(e.g. limb darkening) during light curve analysis and thus we preferred
systems with well determined radiative properties.
 
\subsection{Multiplicity}
 We excluded systems with confirmed third
light in photometry/spectroscopy or known close bright companions
affecting photometric indexes. CD~Tau has a close K-type companion at
a distance of $\sim\!10^{''}$. The light of the companion is present
in the optical light curves analyzed by \cite{rib99} but it was carefully
accounted for in their analysis. The companion is far enough away to not influence
the Tycho or 2MASS magnitudes. Also AI~Phe has a fainter visual companion
(11$^{''}$), the presence of which was accounted for by \cite{kir16} in their
analysis. The case of AI~Phe is actually more complicated as
this system has another, even closer, invisible companion inducing
acceleration on a main binary system (M. Konacki - priv. com.). At
this moment the nature of this companion is uncertain but spectroscopic data
suggests an M type dwarf. In that case its luminosity can be completely
neglected (even in NIR) and we included this system in our sample. 
RR~Lyn is a proposed triple system with a companion of 0.1 $M_\odot$
\citep{kha02}. Even if the companion will be confirmed with future
spectroscopic monitoring at the moment no third light is visible in high
quality light curves \citep[e.g.][]{kha01} and we retained this system
in our sample. 

\subsection{Photometry}
We decided to use homogenous non-saturated optical/infrared photometry from Tycho-2 and the Two Micron All Sky Survey. 

\subsubsection{Optical}
We downloaded the optical $BV$
Tycho-2 photometry \citep{hog00} of the eclipsing binaries from
Vizier \citep{och00}\footnote{http://vizier.u-strasbg.fr:
I/259/tyc2}. $\beta$ Aur, which is by far the brightest star in
our sample, is the only star that lacks Tycho photometry. In this case we used Johnson
photometry from a compilation by \cite{mer91}. Tycho photometry was
subsequently transformed into the Johnson system using the method outlined by
\cite{bes00}. For 6 systems Tycho-2 photometry leads to unexplainable
shifts in the temperatures and surface brightness parameter derived so and
we replaced it by more precise out-of-eclipse optical photometry from
the literature.  

\subsubsection{Near infrared}
We downloaded NIR $JHK_S$
photometry of the Two Micron All Sky Survey (2MASS) \citep{skr06}
from Vizier\footnote{http://vizier.u-strasbg.fr:
II/281/2mass6x}. Magnitudes were converted onto the
Johnson system using equations given in \cite{bes88} and
\cite{car01}\footnote{\texttt{http://www.astro.caltech.edu/$\sim$jmc/2mass/v3/\\transformations/}}.
The transformation equations are as follow: 

\begin{eqnarray*} 
K_J - K_{\rm 2M} & = & 0.037 - 0.017 (J-K)_{\rm 2M} - 0.007 (V-K)_{\rm 2M}  \\
(J-K)_J & = & 1.064 (J-K)_{\rm 2M} +0.006  \\ 
(H-K)_J & = &1.096 (H-K)_{\rm 2M} - 0.027 
\end{eqnarray*}

2MASS photometry of $\beta$~Aur is saturated and we used Johnson $JK$
photometry from a compilation by \cite{duc02}. A lack of good NIR photometry
forced us to remove from the sample the otherwise well suited system $\psi$~Cen.

\begin{deluxetable*}{l@{}ccccccccc}
\tabletypesize{\scriptsize}
\tablecaption{Basic data on the selected detached eclipsing binaries \label{tab:sample}}
\tablewidth{0pt}
\tablehead{
\colhead{Name} & \colhead{Tycho-2} & \colhead{RA$_{2000}$} & \colhead{DEC$_{2000}$} & \colhead{V\tablenotemark{a}} & \colhead{Spectral}&\colhead{Ref.} &\colhead{Orbital}&\colhead{Ref.} & \colhead{Parallax}\\
\colhead{} & \colhead{ID} & \colhead{h:m:s} & \colhead{deg:m:s} & \colhead{mag} & \colhead{Type} & \colhead {SpT} & \colhead{Period (d)} & \colhead{OrP} & \colhead{mas}}
\startdata
YZ Cas& 4307-2167-1& 00:45:39.077& +74:59:17.06& 5.653$\pm$0.015& A2m+F2V& 1  &4.4672235& 36 &10.30$\pm$0.49\\
AI Phe&  8032-0625-1& 01:09:34.195 &$-$46:15:56.09 &8.610$\pm$0.019&F8V+K0IV& 8 & 24.592483 & 40 & 5.94$\pm$0.24\\ 
V505 Per& 3690-0536-1& 02:21:12.964& +54:30:36.28& 6.889$\pm$0.016& F5V+F5V& 2 &4.222020& 2 &15.56$\pm$0.32\\
AL Ari& 0645-1107-1& 02:42:36.341& +12:44:07.77& 9.223$\pm$0.034& F5V+G4V& 3 &3.7474543& 3 &7.11$\pm$0.37\\
V570 Per& 3314-1225-1& 03:09:34.944& +48:37:28.69& 8.091$\pm$0.018& F3V+F5V& 4 &1.9009382& 4 &7.85$\pm$0.26\\
\\
TZ For& 7026-0633-1& 03:14:40.093& $-$35:33:27.60& 6.888$\pm$0.016& F7IV+G8III& 5 &75.66647& 37 &5.379$\pm$0.055\tablenotemark{c}\\
V1229 Tau\tablenotemark{d}& 1800-1622-1& 03:47:29.454& +24:17:18.04& 6.807$\pm$0.017& A0V+Am& 6 &2.46113408& 38 &7.57$\pm$0.40\\
V1094 Tau& 1263-0642-1 &  04:12:03.593& +21:56:50.55 &8.981$\pm$0.031&G0V+G2V& 16 & 8.9885474& 45 & 8.26$\pm$0.25\\
CD Tau& 1291-0292-1& 05:17:31.153& +20:07:54.63& 6.768$\pm$0.016& F6V+F6V& 7 &3.435137& 39 &13.56$\pm$0.38\\
EW Ori& 0104-1206-1 &  05:20:09.147 &+02:02:39.97 &9.902$\pm$0.043\tablenotemark{e}&F8V+G0V&16,47& 6.9368432 & 47 & 5.48$\pm$0.23\\
\\
UX Men& 9378-0190-1& 05:30:03.184& $-$76:14:55.35& 8.251$\pm$0.017& F8V+F8V& 9,\,5 &4.181100& 41 &9.72$\pm$0.21\\
TZ Men& 9496-0590-1& 05:30:13.886& $-$84:47:06.37& 6.186$\pm$0.016& A0V+A8V& 10,\,5 &8.56900& 10 &8.02$\pm$0.49\\
$\beta$ Aur& 2924-2742-1& 05:59:31.723& +44:56:50.76& 1.900$\pm$0.020\tablenotemark{f}& A1mIV+A1mIV& 11,57 &3.960047& 42 &40.21$\pm$0.23\tablenotemark{b}\\
RR Lyn& 3772-2770-1& 06:26:25.836& +56:17:06.35& 5.558$\pm$0.015& A6mIV+F0V& 12 &9.945074& 12 &13.34$\pm$0.60\tablenotemark{b}\\
WW Aur& 2426-0345-1& 06:32:27.185& +32:27:17.63& 5.832$\pm$0.016& A4m+A5m& 13 &2.5250194& 43 &11.03$\pm$0.50\\
\\
HD 71636& 2489-1972-1& 08:29:56.311& +37:04:15.48& 7.903$\pm$0.018& F2V+F5V& 14 &5.013292& 14 &8.40$\pm$0.40\\
VZ Hya&  4874-0811-1& 08:31:41.413 &$-$06:19:07.56 & 8.953$\pm$0.027\tablenotemark{g}&F3V+F5V& 21 & 2.9043002 & 51 & 6.94$\pm$0.24\\
KX Cnc& 2484-0592-1& 08:42:46.211& +31:51:45.37& 7.192$\pm$0.017& F9V+F9V& 15 &31.2197874& 44 &20.54$\pm$0.38\\
PT Vel& 7690-2859-1& 09:10:57.720& $-$43:16:02.93& 7.027$\pm$0.016& A0V+F0& 17 &1.802008& 17 &6.15$\pm$0.45\\
KW Hya& 4891-1371-1& 09:12:26.044& $-$07:06:35.38& 6.100$\pm$0.016& A5m+F0V& 18 &7.750469& 46 &11.53$\pm$0.42\tablenotemark{b}\\
\\
RZ Cha& 9422-0104-1& 10:42:24.104& $-$82:02:14.19& 8.091$\pm$0.018& F5V+F5V& 20  &2.832084& 48 &5.68$\pm$0.26\\
FM Leo& 0263-0727-1& 11:12:45.095& +00:20:52.83& 8.460$\pm$0.021& F7V+F7V& 21 &6.728606& 49 &7.00$\pm$0.32\\
GG Lup& 7826-3079-1& 15:18:56.376& $-$40:47:17.60& 5.603$\pm$0.015& B7V+B9V& 23 &1.8495927& 50 &5.96$\pm$0.30\tablenotemark{b}\\
V335 Ser & 0353-0301-1&  15:59:05.756 &+00:35:44.55& 7.490$\pm$0.017& A1V+A3V & 19 &  3.4498837 & 19 & 4.74$\pm$0.30\\
WZ Oph& 0977-0216-1& 17:06:39.042& +07:46:57.78& 9.126$\pm$0.024& F8V+F8V& 24,\,25 &4.183507& 51 &6.61$\pm$0.24\\
\\
FL Lyr&3542-1492-1 & 19:12:04.862& +46:19:26.86& 9.366$\pm$0.026& F8V+G8V& 27 &2.1781542& 27 &7.25$\pm$0.22\\
UZ Dra & 4444-1595-1 & 19:25:55.045 &+68:56:07.15 & 9.601$\pm$0.028& F7V+G0V &22 & 3.261302 & 52 & 5.21$\pm$0.25\\
V4089 Sgr& 7936-2270-1& 19:34:08.486& $-$40:02:04.70& 5.907$\pm$0.016& A2IV+A7V& 28,\,29 &4.6270956& 29 &6.75$\pm$0.49\\
V1143 Cyg& 3938-1983-1& 19:38:41.184& +54:58:25.65& 5.889$\pm$0.015& F5V+F5V& 30,\,31 &7.640742& 54 &24.75$\pm$0.35\\
MY Cyg & 2680-1529-1 & 20:20:03.390 &+33:56:35.02&  8.341$\pm$0.019&F0m+F0m &32 & 4.0051870& 55 & 3.95$\pm$0.24\\
\\
EI Cep& 4599-0082-1& 21:28:28.206& +76:24:12.59& 7.600$\pm$0.017& F3V+F1V& 33 &8.4393522&  33 &5.07$\pm$0.24\\
VZ Cep& 4470-1334-1&  21:50:11.135 &+71:26:38.30& 9.717$\pm$0.009\tablenotemark{h}& F3V+G4V & 26 & 1.1833638 & 26 & 3.88$\pm$0.35\\
LL Aqr& 5236-0883-1& 22:34:42.152& $-$03:35:58.17& 9.243$\pm$0.037\tablenotemark{i}& F8V+G2V& 34 &20.178321& 56  &7.75$\pm$0.27\\
EF Aqr&  5248-1030-1&  23:01:19.088 & -06:26:15.35 & 9.885$\pm$0.022\tablenotemark{j} & F8V+G8V & 15 & 2.8535721 & 53 & 5.06$\pm$0.50\\
V821 Cas& 4001-1445-1& 23:58:49.175& +53:40:19.81& 8.286$\pm$0.017& A1V+A4& 35 &1.7697397& 35 &3.61$\pm$0.30
\enddata
\tablecomments{{\bf Ref.} to Spectral Type (SpT) and/or Orbital Period (OrP): 1 - \cite{pav14}; 2 - \cite{tom08a}; 3 - \cite{kon17}; 4 - \cite{tom08b}; 5 - \cite{tor10}; 6 - \cite{abt78}; 7 - \cite{pop71}; 8 - \cite{and88}; 9 - \cite{hou75}; 10 - \cite{and87a}; 11 - \cite{noj94}; 12 - \cite{kha01}; 13 - \cite{kiy75}; 14 - \cite{hen06}; 15 - this work; 16 - \cite{nes95}; 17 - \cite{bak08}; 18 - \cite{and91}; 19 - \cite{slac12}; 20 - \cite{pop66}; 21 - \cite{hou99}; 22 - \cite{slac89}; 23 - \cite{and93}; 24 - \cite{pop65}; 25 - \cite{bat78}; 26 - \cite{tor09};  27 - \cite{pop86}; 28 - \cite{hou78}; 29 - \cite{ver15}; 30 - \cite{hil75}; 31 - \cite{and87b}; 32 - \cite{mal93}; 33 - \cite{tor00}; 34 - \cite{gri13}; 35 - \cite{cak09}; 36 - \cite{lac81}; 37 - \cite{gal16}; 38 - \cite{tre16}; 39 - \cite{rib99}; 40 - \cite{kir16}; 41 - \cite{cla76}; 42 - \cite{sou07}; 43 - \cite{sou05}; 44 - \cite{sow12};  45 - \cite{max15}; 46 - \cite{and84}; 47 - \cite{cla10}; 48 - \cite{jor75}; 49 - \cite{rat10}; 50 - \cite{bud15}; 51 - \cite{cla08a}; 52 - \cite{gul86}; 53 - \cite{vos12}; 54 - \cite{gim85}; 55 - \cite{tuc09}; 56 - \cite{sou13}; 57 - \cite{lyu96}\vspace{3pt}}
\tablenotetext{a}{Tycho-2 V$_T$ magnitudes from  \cite{hog00} converted onto Johnson V magnitudes using transformation given by \cite{bes00}}
\tablenotetext{b}{Hipparcos parallax \citep{vLe07}}
\tablenotetext{c}{Orbital parallax from \cite{gal16}, Hipparcos value $\varpi=5.75\pm0.51$ mas, Gaia value $\varpi=5.44\pm0.25$ mas}
\tablenotetext{d}{HD 23642, in the Pleiades cluster}
\tablenotetext{e}{\cite{cla10}}
\tablenotetext{f}{\cite{mer91}}
\tablenotetext{g}{\cite{cla08a}}
\tablenotetext{h}{\cite{slac02}}
\tablenotetext{i}{\cite{gra16}}
\tablenotetext{j}{\cite{vos12}}
\end{deluxetable*}


\section{Method}\label{method}
\subsection{The Wilson-Devinney model of the systems}
\label{sec:wd}
For the purpose of obtaining homogenous parameters for the 
eclipsing binary sample we decided to create a model of each
system. The models were built using the Wilson-Devinney code version
2007 \citep{wil71,wil79,wil90,van07} while parameters of the models were
based on solutions published in the literature. None of the eclipsing
binaries in our sample has infrared $J,H,K$ light curves suitable for
deriving direct light ratios in those bands. Thus, in order to calculate
intrinsic infrared colors of the components of each system we employed
eclipsing binary models based on optical light curves and we extrapolated
them into infrared. Such an approach may introduce some bias which will
be discussed later in this paper. All models were checked for internal
consistency of the parameters and it turned out that in many cases they had
to be fine-tuned. In particular, the temperature ratio and the absolute 
temperature scale, being crucial for precise prediction of infrared 
light ratios, were inspected carefully.

The procedure was as follows. For each system we collected orbital
and photometric parameters from the most recent publications. The
input parameters were the radial velocity semi-amplitudes $K_{1,2}$,
the orbital period $P$, three parameters describing the position of the orbit
(the orbital inclination $i$, the eccentricity $e$ and the longitude of
periastron $\omega$), the photometric relative radii $r_{1,2}$ and the
effective temperatures $T_{1,2}$. Those parameters were transformed
into the semi-major axis of the system $a$, the mass ratio $q$ and into
dimensionless Roche potentials $\Omega_{1,2}$ using equations given in
\cite{tor10} and \cite{wil79}, i.e. parameters directly fitted or used
within the WD program. We usually fixed the temperature of the primary star
$T_1$ and then, using published light ratios in different photometric
bands, we adjusted the temperature of the companion $T_2$. In some cases
however we also re-derived $T_1$ as it is described later. The rotation
parameter $F_{1,2}$ was kept to 1 (synchronous rotation), unless there
was a direct spectroscopic determination of $F$ significantly different
from unity. The albedo $A$ and the gravity brightening $g$ were set
in a standard way for a convective atmosphere cooler than 7200~K and
radiative ones for a hotter surface temperature. This was done only for
the sake of consistency because the two parameters have negligible effect on
the light ratios. The input and derived parameters used to create 
the appropriate WD models are listed in Table~\ref{tab:par}. 

\subsection{Correction of 2MASS magnitudes taken during eclipses}
KX~Cnc, GG~Lup and WW~Oph have 2MASS observations taken during the secondary
eclipses. To account for a light lost during minima we used our models to
calculate the appropriate corrections. The corrections are $\Delta J=-0.333$
mag, $\Delta H=-0.331$ mag and $\Delta K=-0.331$ mag for KX~Cnc, $\Delta
J=-0.277$ mag, $\Delta H=-0.281$ mag and $\Delta K=-0.285$ mag for 
GG~Lup, $\Delta J=-0.392$ mag, $\Delta H=-0.390$ mag and $\Delta K=-0.390$
mag for WZ~Oph. For GG~Lup we accounted also for the apsidal
motion which shifts the position of the eclipses \citep{wol05}.

\begin{turnpage}
\begin{deluxetable*}{@{}lcccccccccccccc@{}}
\tabletypesize{\scriptsize}
\tablecaption{Parameters of the Wilson-Devinney models \label{tab:par}}
\tablewidth{0pt}
\tablehead{
\colhead{}& \multicolumn{9}{c}{Input parameters}&\colhead{Reference}&\multicolumn{4}{c}{Model parameters}\\
\colhead{Eclipsing} & \multicolumn{2}{c}{RV semiamplitude}& \multicolumn{3}{c}{Orientation of the orbit}& \multicolumn{2}{c}{Fractional radius}& \multicolumn{2}{c}{Effective temperature}& \colhead{} & \colhead{Semimajor} & \colhead{Mass} & \colhead{$\Omega_1$} &\colhead{$\Omega_2$}\\
\colhead{binary} & \colhead{$K_1 ($km s$^{-1}$)}& \colhead{$K_2 ($km s$^{-1}$)} & \colhead{$e$} &\colhead{$\omega$ (rad)}& \colhead{$i$ (deg)} & \colhead{$r_1$} & \colhead{$r_2$} & \colhead{$T_1 ($K)} & \colhead{$T_2 ($K)}& \colhead{} & \colhead{axis$(R_\odot)$} & \colhead{ratio}&\colhead{}&\colhead{}}
\startdata
   YZ Cas & 73.05(19) & 124.78(27)& 0.0 & 0.0 & 88.33(7) & 0.14456(56) & 0.07622(33)& 9520(120)&6880(240) & 2 & 17.4764 & 0.5854 & 7.5141& 8.8912 \\
  AI Phe\tablenotemark{d}  &  51.128(28) & 49.120(19) & 0.187(3)& 1.933(6)& 88.55(5) & 0.03845(35) & 0.06070(27) &  6175(150) & 5140(120) & 3,4,5,1&47.8850&1.0409 &27.290&18.361\\
 V505 Per & 89.01(8) & 90.28(9) & 0.0 & 0.0 &87.95(4) & 0.0860(9) & 0.0846(9) & 6512(21)&6460(30) &6 & 14.9724 & 0.9859 & 12.618 & 12.665  \\
   AL Ari &   76.98(13)      & 98.38(21) &  0.051(3)& 1.20(2) & 89.48(6)& 0.1060(4)& 0.0696(3)& 6300(80)& 5412(80)& 7 &12.9738 & 0.7825&10.269 & 12.411 \\
 V570 Per\tablenotemark{a} &  114.09(27)  & 122.48(28) & 0.0 &0.0 &77.4(3) & 0.1675(31) & 0.1526(19) & 6842(30) & 6562(30) & 8,1 & 9.10738 & 0.9315 & 6.9200 & 7.1538  \\
 \\
   TZ For\tablenotemark{b} &  40.868(11) & 38.900(22) & 0.0&0.0 & 85.66(4)& 0.03320(70) & 0.06972(92)& 6350(70) & 4930(30) & 9,10 & 119.650 & 1.0506 & 31.416 & 16.046 \\
V1229 Tau & 99.02(27) & 140.86(36) & 0.0 & 0.0 & 78.2(1) & 0.1450(23) & 0.1262(37) & 9950(300) & 7640(300) & 11,12 & 11.9214 & 0.7030 & 7.6115 & 6.7280 \\
V1094 Tau\tablenotemark{c} & 65.38(7) & 70.83(12) & 0.2677(4) & 5.822(3) & 88.21(1) & 0.06050(24) & 0.04744(29) & 5850(100) & 5720(100) & 13,1 & 23.3292 & 0.9231 & 17.792 & 20.863 \\
   CD Tau &  96.8(5) & 102.1(5) & 0.0 & 0.0& 87.7(3)& 0.1330(10) & 0.1172(13) & 6200(50) & 6194(50) & 14 & 13.5172 & 0.9481 & 8.4785 & 9.1246 \\
   EW Ori\tablenotemark{c} & 72.48(21) & 75.45(25) & 0.076(2) & 5.40(2) & 89.86(9) & 0.05786(18) & 0.05434(18) & 6070(95) &  5875(95) & 15,16,17,1 & 20.2258 & 0.9607 & 18.325 & 18.783 \\
 \\
   UX Men\tablenotemark{c} & 87.36(17) &90.08(14) &0.003(3) & 1.3(6) & 89.6(1) & 0.0918(9) & 0.0868(9) & 6200(100) & 6127(100) & 18,3,1 & 14.6652 & 0.9682 & 11.871& 12.196 \\ 
   TZ Men & 62.15(12) & 102.82(45) & 0.035(7)& 4.75(2)& 88.7(1)& 0.0722(7) & 0.0513(5) & 10400(500) & 7240(300) & 19 & 27.9328 & 0.6045 & 14.480 & 13.018 \\
 $\beta$ Aur\tablenotemark{c} & 108.04(10) & 110.93(10) & 0.0018(4) & 1.579(5) & 76.8(1)& 0.15694(81) & 0.14595(82) & 9350(200)& 9297(200) & 20,21,22,23,1& 17.6051 &0.9739 &7.3643 & 7.7040  \\
   RR Lyn & 65.65(6) & 83.92(17) & 0.079(1)& 3.14(1)&87.5(1) & 0.0878(5) & 0.0541(11) & 7570(120) & 6980(100) & 24,25 & 29.340 & 0.7823& 12.244 & 15.657  \\
   WW Aur\tablenotemark{d} & 116.81(23) & 126.49(28) & 0.0 & 0.0& 87.55(4)& 0.1586(9) & 0.1515(9) & 8180(260) & 7872(250) & 26,1 & 12.1546 & 0.9235 & 7.2450 & 7.1487  \\
 \\  
  HD 71636\tablenotemark{a} & 80.30(18) & 94.45(19) & 0.0& 0.0& 85.63(2)& 0.0904(5) & 0.0784(5) & 6950(140) & 6440(140) & 27,1 & 17.3682 & 0.8502 & 11.917 & 11.923 \\
  VZ Hya & 94.92(19) & 105.31(34) & 0.0 & 0.0 & 88.88(9) & 0.1143(4) & 0.0968(6) & 6645(150) & 6300(150) & 28 & 11.4972 & 0.9013 & 9.6587 & 10.367 \\
   KX Cnc\tablenotemark{a,d} & 50.039(65) & 50.503(65) & 0.4667(1) &1.113(1) & 89.83(1) & 0.01940(4) & 0.01913(5) & 6050(110) & 5995(110) & 29,1 & 54.8787 & 0.9908 & 53.405 & 53.674  \\
   PT Vel &   117.2(2) & 158.5(5) & 0.127(6)& 5.06(1)&88.2(5) & 0.215(2) & 0.160(2) & 9250(150) & 7650(155) & 30 & 9.7457 & 0.7394 & 5.5259 & 5.9125 \\
   KW Hya\tablenotemark{c} & 70.12(21) & 93.17(79) &0.0945(1) & 3.929(2)& 87.65(4)& 0.0853(5) & 0.0594(8) & 8000(200) & 6960(210) & 31,1 & 24.9247 & 0.7526 &12.559 & 13.908 \\
 \\  
   RZ Cha\tablenotemark{d} & 108.2(6) & 107.6(9) & 0.0& 0.0& 82.89(7)& 0.1777(20) & 0.1893(40) & 6580(150) & 6530(150) & 32,33,1 & 12.1746 & 1.0056 & 6.6545 & 6.3340  \\
   FM Leo & 76.62(27) & 78.46(28) & 0.0 &0.0 &87.98(6) & 0.0798(21) & 0.0732(24) & 6316(240) & 6190(210) & 34 & 20.6392& 0.9765& 13.505& 14.354 \\
   GG Lup\tablenotemark{c} & 125.1(5)& 203.4(8) & 0.154(5)& 2.351(3)& 86.8(1)& 0.2003(18)& 0.1456(14) & 14750(450) & 11200(500) & 35,36,1 & 11.8871& 0.6150& 5.7696 & 5.6076  \\
   V335 Ser & 106.57(12) & 120.07(38) & 0.141(2) & 1.139(7) & 87.2(2) & 0.1325(17) & 0.1131(22) &  9020(150) & 8500(150) & 37 & 15.3191 & 0.8876 & 8.5918 & 9.0764 \\
   WZ Oph\tablenotemark{d} & 88.77(19) & 89.26(24) &0.0 & 0.0&89.1(1) & 0.0952(8) & 0.0964(8) & 6232(100) & 6212(100) & 28,1 &14.7240 & 0.9945 & 11.505& 11.326 \\
 \\
   FL Lyr\tablenotemark{c} & 93.5(5)& 118.9(7)& 0.0& 0.0& 86.3(4)& 0.140(3) & 0.105(3) & 6150(120)& 5270(110)& 15,1& 9.1640 & 0.7864 & 7.9411& 8.6028  \\
   UZ Dra\tablenotemark{d} & 93.52(35) & 101.55(43) & 0.0 & 0.0 & 89.1(2) & 0.103(2) & 0.091(2) & 6450(120)& 6170(120) & 38,39,1 & 12.5768 & 0.9209 & 10.637 & 11.165 \\
V4089 Sgr\tablenotemark{a} & 78.48(18)& 126.20(24)& 0.0&0.0 & 83.48(6)&0.2104(9)&0.0852(3) & 8433(100)& 7361(105) & 40,1 &18.8426 & 0.6219& 5.3991& 8.4925 \\
V1143 Cyg & 88.02(5)& 89.97(10) &0.5378(3) & 0.860(1)& 87.0(1)&0.059(1) & 0.058(1)& 6450(100) & 6400(100) & 41,42 & 22.6950& 0.9783& 19.069& 19.045  \\
MY Cyg &101.9(8) &103.3(6) &0.010(1)&1.21(4)&88.58(2)&0.138(3)&0.134(3)&7050(200)&7000(200)&43,44,45&16.2482&0.9867&8.2580&8.3912\\
 \\
   EI Cep\tablenotemark{c} & 76.84(13)& 81.02(13)& 0.0& 0.0& 87.23(9)&0.1099(20) & 0.0884(18)& 6750(120)& 6977(120) & 46,1&26.3649 & 0.9484 & 10.056 & 11.760  \\
  VZ Cep\tablenotemark{c} & 118.88(22) & 150.48(67) & 0.0 & 0.0 & 79.97(5) & 0.2398(17) & 0.1630(61) & 6690(160) & 5705(120) & 47,1 & 6.3985 & 0.7900 & 4.9955 & 5.9680 \\
   LL Aqr & 49.948(13) & 57.736(14)& 0.3165(1) & 2.714(1) & 89.55(3)  & 0.03246(5)& 0.02459(8)&6080(50) & 5705(60) & 48,49 & 40.7438& 0.8651 &32.074 &  36.708\\
   EF Aqr\tablenotemark{a} &84.175(66) &110.66(24)&0.0&0.0&88.45(8)&0.1222(9)&0.0876(7)&6150(65) &5185(110)& 50,1 & 10.9940 & 0.7607 & 8.9529 & 9.8076\\
 V821 Cas\tablenotemark{a} & 120.8(1.7)& 152.4(2.0) & 0.127(7)& 2.71(7)&82.6(1) & 0.2434(13)& 0.1466(17) & 9400(400) & 8600(400) & 51,1 & 9.5600 & 0.7927 & 5.0538 &  6.6690
\enddata 
\tablecomments{Reference: 1 - this paper; 2 - \cite{pav14}; 3 - \cite{hel09}; 4 - \cite{and88}; 5 - \cite{kir16}; 6 - \cite{tom08a}; 7 - \cite{kon17}; 8 - \cite{tom08b}; 9 - \cite{and91b}; 10 - \cite{gal16}; 11 - \cite{tre16}; 12 - \cite{gro07}; 13 - \cite{max15}; 14 - \cite{rib99};  15 - \cite{pop86}; 16 - \cite{imb02}; 17 - \cite{cla10}; 18 - \cite{and89}; 19 - \cite{and87a}; 20 - \cite{smi48}; 21 - \cite{beh11}; 22 - \cite{noj94}; 23 - \cite{sou07}; 24 - \cite{tom06}; 25 - \cite{kha01}; 26 - \cite{sou05}; 27 - \cite{hen06}; 28 - \cite{cla08b}; 29 - \cite{sow12};  30 - \cite{bak08}; 31 - \cite{and84};  32 - \cite{and75b}; 33 - \cite{giu80};  34 - \cite{rat10};  35 - \cite{and93}; 36 - \cite{bud15}; 37 - \cite{slac12}; 38 - \cite{imb86}; 39 - \cite{slac89}; 40 - \cite{ver15}; 41 - \cite{alb07}; 42 - \cite{and87b};  43 - \cite{pop71}; 44 - \cite{tuc09}; 45 - \cite{tor10}; 46 - \cite{tor00}; 47 - \cite{tor09}; 48 - \cite{sou13}; 49 - \cite{gra16}; 50 - \cite{vos12}; 51 - \cite{cak09}}
\tablenotetext{a}{We recalculated radial velocity semiamplitudes - see Section \ref{sec:rv}.}
\tablenotetext{b}{We set rotation parameter $F_1=18.0$}
\tablenotetext{c}{We adjusted temperature $T_2$.}
\tablenotetext{d}{We recalculated both temperatures - see Section \ref{sec:tempred}}
\end{deluxetable*}
\end{turnpage}

\subsection{Temperature and Reddening} 
\label{sec:tempred}
In some individual cases, described in the Appendix the temperature 
$T_1$ or/and color excess E(B$-$V) were adjusted in order to obtain agreement between
intrinsic colors and temperatures. Reddenings to each object were taken
from the literature, if available, and also derived independently
using the extinction maps by \cite{sch98} following the prescription given in
\cite{suc15}. Dereddened magnitudes and colors were calculated using
the mean Galactic interstellar extinction curve from \cite{fit07} assuming
$R_V=3.1$.  To re-derive temperatures we used a number of calibrations
given below: 
\begin{itemize}
\item{$b-y$: \cite{hol07}, \cite{ram05}, \cite{alo96}, \cite{nap93}.}
\item{$B\!-\!V$: \cite{cas10}, \cite{gon09}, \cite{ram05}, \cite{flo96}.}
\item{$V\!-\!J$: \cite{cas10}, \cite{gon09}.}
\item{$V\!-\!K$: \cite{wor11}, \cite{cas10}, \cite{gon09}, \cite{mas06}, \cite{ram05}, \cite{hou00}, \cite{alo96}.}
\end{itemize}

\subsection{Radial velocity semi-amplitudes}
\label{sec:rv}
Usually we assumed radial velocity semi-amplitudes from the literature. When
two or more orbital solutions were published based on different radial
velocity sets and having uncertainties of the same order of magnitude,
we used the weighted mean to derive the final parameters, i.e. AI~Phe, 
EW~Ori, UX~Men, $\beta$~Aur, GG~Lup, UZ~Dra, and V1143~Cyg. However, in few
cases we redetermined the spectroscopic orbits from source data in order
to derive directly $K_{1,2}$ or to check the consistency of the orbital
parameters and their errors. The spectroscopic orbits were derived with
the Wilson-Devinney code taking into account the full model of a system and
all proximity effects. A set of numerical constants used to change from SI
units into astrophysical units were chosen after \cite{tor10}. Individual
cases are described in the Appendix.

\begin{turnpage}
\begin{deluxetable*}{@{}lccccccccccccccccc@{}}
\tabletypesize{\scriptsize}
\tablecaption{Photometric and physical parameters used to derive individual angular diameters and colors. \label{tab:fot}}
\tablewidth{0pt}
\tablehead{
\colhead{Eclipsing} & \colhead{E$(B-V)$ } & \colhead{Ref.} & \multicolumn{2}{c}{Radius ($R_\odot$)}
&\multicolumn{2}{c}{Distance (pc)} & \colhead{$\sigma$} &\multicolumn{5}{c}{Unreddened Johnson photometry\tablenotemark{a} (mag)} & \multicolumn{5}{c}{Light ratio $L_2/L_1$\tablenotemark{b}}\\ 
\colhead{binary} & \colhead{mag} & \colhead{} & \colhead{$R_1$} & \colhead{$R_2$} & \colhead{Geom.} & \colhead{Photom.\tablenotemark{c}} & \colhead{} & \colhead{$B_0$}
& \colhead{$V_0$} &\colhead{$J_0$} & \colhead{$H_0$} & \colhead{$K_0$} & \colhead{$B$} & \colhead{$V$} & \colhead{$J$} & \colhead{$H$} &\colhead{$K$} }
\startdata
    YZ Cas &  0.015(10)&1 & 2.526(11) & 1.332(6)& 97.1(4.6) & 99.2(4.0) & 0.35 &5.657(48) &  5.607(34) &  5.616(21) & 5.652(42) & 5.635(22) & 0.0610 & 0.0882 & 0.1682 & 0.2004 & 0.2046 \\
    AI Phe &  0.012(10)& 2,1& 1.841(17) & 2.907(13)& 168.4(6.8) & 167.9(6.7)&0.15 & 9.212(52) &  8.573(36) &  7.345(25) & 6.930(38) & 6.832(27) & 0.7382 & 1.0057 & 1.6394 & 1.9685 & 1.9828 \\
  V505 Per &  0.003(5)&1 & 1.288(14) & 1.267(14)& 64.3(1.3) & 60.7(9) & 2.23&7.287(34) &  6.880(22) &  6.117(70) & 5.791(40) & 5.794(21) & 0.9244 & 0.9348 & 0.9522 & 0.9584 & 0.9588 \\
    AL Ari &  0.012(10)& 3& 1.375(6) & 0.903(4)& 140.6(7.3) & 137.0(4.0) & 0.45&9.696(69) &  9.186(46) &  8.235(23) & 7.933(23) & 7.905(27) & 0.1646 & 0.2100 & 0.3087 & 0.3571 & 0.3597 \\
  V570 Per &  0.070(30)&1 & 1.525(30) & 1.390(19)& 127.4(4.2) & 118.6(5.4) & 1.29&8.270(126) &  7.875(94) &  7.156(35) & 6.921(23) & 6.888(22) & 0.6567 & 0.6950 & 0.7645 & 0.7916 & 0.7926 \\
  \\
    TZ For &  0.015(5)& 4& 3.972(84) & 8.34(11)& 185.9(1.9) & 185.2(3.8) &0.17 &7.569(34) &  6.842(22) &  5.530(21) & 5.124(26) & 5.007(30) & 0.7888 & 1.2341 & 2.4644 & 3.1856 & 3.2193 \\
 V1229 Tau &  0.020(10)& 5,6& 1.729(27) & 1.505(45)& 132.1(7.0) & 133.1(7.5) &0.10 & 6.784(50) &  6.745(35) &  6.663(25) & 6.644(29) & 6.637(25) & 0.2628 & 0.3385 & 0.5223 & 0.5800 & 0.5917 \\
 V1094 Tau &  0.026(10)& 7 & 1.411(6) & 1.107(7)& 121.1(3.7) & 118.0(3.8) &0.58& 9.575(66) &  8.901(44) &  7.814(23) & 7.520(46) & 7.437(22) & 0.5318 & 0.5524 & 0.5851 & 0.5980 & 0.5986 \\
    CD Tau &  0.005(5)& 8,1 & 1.798(17) & 1.584(20)& 73.7(2.1) & 68.6(1.2) &2.22& 7.231(34) &  6.753(22) &  5.894(22) & 5.671(34) & 5.612(30) & 0.7702 & 0.7724 & 0.7749 & 0.7764 & 0.7760 \\
    EW Ori &  0.026(14)&9 & 1.170(5) & 1.099(5)& 182.5(7.7) & 173.8(6.3) &0.88& 10.407(94) &  9.822(61) &  8.837(26) & 8.598(69) & 8.513(22) & 0.7209 & 0.7595 & 0.8223 & 0.8474 & 0.8486 \\
    \\
    UX Men &  0.027(10)& 10& 1.346(13) & 1.273(13)& 102.9(2.2) & 100.6(2.9) &0.62& 8.692(50) &  8.168(35) &  7.222(29) & 6.966(30) & 6.931(25) & 0.8318 & 0.8476 & 0.8721 & 0.8818 & 0.8822 \\
    TZ Men &  0.000(5)& 11& 2.017(20) & 1.433(15)& 124.7(7.6) & 117.7(8.2) &0.63& 6.166(34) &  6.186(22) &  6.180(31) & 6.128(44) & 6.153(27) & 0.1150 & 0.1623 & 0.3002 & 0.3508 & 0.3607 \\
  beta Aur &  0.000(3)& 1& 2.763(15) & 2.569(15)& 24.9(1) & 25.0(9) &0.26& 1.930(34) &  1.900(22) &  1.869(42) & --- & 1.859(41) & 0.8462 & 0.8524 & 0.8585 & 0.8597 & 0.8603 \\
    RR Lyn &  0.007(5)& 1& 2.576(20) & 1.587(30)& 75.0(3.4) & 72.6(2.0) &0.50& 5.764(33) &  5.536(22) &  --- & 5.073(23) & 5.021(17) & 0.2343 & 0.2645 & 0.3298 & 0.3534 & 0.3541 \\
    WW Aur &  0.008(5)&1 & 1.928(11) & 1.841(11)& 90.7(4.1) & 85.8(4.0) &1.02& 5.976(34) &  5.807(22) &  5.533(22) & 5.505(29) & 5.513(22) & 0.7600 & 0.7953 & 0.8607 & 0.8753 & 0.8789 \\
    \\
   HD71636 &  0.020(10)&12 & 1.570(9) & 1.362(7)& 119.0(5.7) & 118.6(4.1) &0.07& 8.223(51) &  7.841(36) &  7.104(22) & 6.917(24) & 6.907(35) & 0.4910 & 0.5442 & 0.6474 & 0.6886 & 0.6911 \\
    VZ Hya &  0.027(20)& 13& 1.314(5) & 1.113(7)& 144.1(5.0) & 146.0(7.3) &0.23& 9.307(93) &  8.870(67) &  8.105(29) & 7.844(28) & 7.801(18) & 0.5254 & 0.5683 & 0.6439 & 0.6747 & 0.6766 \\
    KX Cnc &  0.001(5)& 1& 1.065(2) & 1.050(3)& 48.7(9) & 49.0(1.4) &0.19& 7.766(35) &  7.189(23) &  6.223(32) & 5.949(30) & 5.905(31) & 0.9196 & 0.9330 & 0.9537 & 0.9616 & 0.9621 \\
    PT Vel &  0.005(5)&14 & 2.095(20) & 1.559(20)& 163(12) & 164.8(4.9) &0.19& 7.062(34) &  7.012(22) &  6.902(32) & 6.861(32) & 6.871(30) & 0.2395 & 0.2949 & 0.4199 & 0.4580 & 0.4641 \\
    KW Hya &  0.006(6)& 1& 2.126(15) & 1.480(22)& 86.7(3.2) & 86.5(3.6) &0.05& 6.308(37) &  6.081(24) &  5.694(24) & 5.642(44) & 5.574(22) & 0.2249 & 0.2699 & 0.3816 & 0.4221 & 0.4255 \\
    \\
    RZ Cha &  0.038(20)& 1& 2.163(20) & 2.305(20)& 176.1(8.1) & 179.4(8.0) &0.37& 8.384(87) &  7.974(64) &  7.148(35) & 6.926(40) & 6.919(39) & 1.0879 & 1.0988 & 1.1177 & 1.1241 & 1.1250 \\
    FM Leo &  0.019(10)& 1& 1.648(43) & 1.511(49)& 142.9(6.5) & 139.6(8.8) &0.27& 8.884(53) &  8.401(37) &  7.554(23) & 7.324(56) & 7.229(24) & 0.7441 & 0.7681 & 0.8062 & 0.8216 & 0.8222 \\
    GG Lup &  0.027(10)& 15& 2.381(22) & 1.732(17)& 167.8(8.4) & 147.3(9.5) &1.61& 5.386(48) &  5.520(34) &  5.861(32) & 5.980(38) & 5.961(31) & 0.3026 & 0.3236 & 0.3774 & 0.3875 & 0.3964 \\
  V335 Ser &  0.068(8)&16 & 2.030(26) & 1.733(34)& 211(13) & 195.2(6.0) &1.15& 7.352(43) &  7.280(30) &  7.127(23) & 7.105(36) & 7.080(22) & 0.5730 & 0.6109 & 0.6720 & 0.6870 & 0.6901 \\
    WZ Oph &  0.030(16)& 1& 1.402(12) & 1.420(12)& 151.3(5.5) & 164.9(5.7) &1.72& 9.543(76) &  9.033(55) &  8.208(34) & 7.972(37) & 7.894(31) & 1.0068 & 1.0115 & 1.0188 & 1.0214 & 1.0217 \\
    \\
    FL Lyr &  0.010(7)& 1& 1.283(30) & 0.962(30)& 137.9(4.2) & 131.9(5.2) &0.83& 9.875(52) &  9.335(34) &  8.285(28) & 7.983(36) & 7.917(21) & 0.2065 & 0.2646 & 0.3969 & 0.4623 & 0.4661 \\
    UZ Dra &  0.012(7)& 17& 1.295(25) & 1.144(25)& 191.9(9.2) & 189.7(6.6) &0.20& 10.036(54) &  9.564(35) &  8.653(22) & 8.423(22) & 8.393(20) & 0.5991 & 0.6411 & 0.7124 & 0.7419 & 0.7431 \\
 V4089 Sgr &  0.027(15)& 1& 3.964(20) & 1.605(7)& 148(11) & 145.5(4.3) &0.19& 5.889(67) &  5.824(49) &  5.710(25) & 5.627(40) & 5.623(25) & 0.0776 & 0.0953 & 0.1327 & 0.1452 & 0.1466 \\
 V1143 Cyg &  0.000(5)& 18,1& 1.339(23) & 1.316(23)& 40.4(6) & 39.2(1.1) &1.02& 6.347(34) &  5.889(22) &  5.024(21) & 4.845(22) & 4.798(21) & 0.9233 & 0.9339 & 0.9512 & 0.9577 & 0.9581 \\
    MY Cyg &  0.048(30)&19 & 2.242(50) & 2.178(50)& 253(15) & 229(14) &1.17& 8.520(126) &  8.193(94) &  7.702(51) & 7.563(46) & 7.538(26) & 0.9075 & 0.9158 & 0.9312 & 0.9360 & 0.9363 \\
    \\
    EI Cep &  0.007(5)& 20& 2.898(48) & 2.331(44)& 197.2(9.3) & 194.0(5.8) &0.30& 7.956(35) &  7.578(23) &  6.879(21) & 6.738(37) & 6.696(18) & 0.7702 & 0.7405 & 0.6891 & 0.6742 & 0.6723 \\
    VZ Cep &  0.044(10)&21 & 1.534(12) & 1.043(39)& 258(23) & 211.1(8.5) &2.07& 10.025(46) &  9.581(32) &  8.796(23) & 8.621(33) & 8.606(22) & 0.1764 & 0.2256 & 0.3308 & 0.3830 & 0.3851 \\
    LL Aqr &  0.018(14)& 22& 1.323(6) & 1.002(5)& 129.0(4.5) & 132.4(3.6) &0.57& 9.748(86) &  9.187(57) &  8.180(26) & 7.864(37) & 7.836(24) & 0.3848 & 0.4268 & 0.4998 & 0.5306 & 0.5320 \\
    EF Aqr &  0.025(15)& 23& 1.343(10) & 0.963(8)& 198(20) & 169.4(6.0) &1.57& 10.385(77) &  9.808(51) &  8.849(27) & 8.511(29) & 8.496(24) & 0.1678 & 0.2207 & 0.3480 & 0.4141 & 0.4174 \\
  V821 Cas &  0.060(30)&1 & 2.327(29) & 1.401(33)& 277(23) & 306(24) &0.91& 8.158(125) &  8.101(94) &  7.976(29) & 7.994(32) & 7.975(29) & 0.2554 & 0.2829 & 0.3231 & 0.3333 & 0.3354 
\enddata                                                                 
\tablecomments{References to reddening: 1 - this work; 2 - \cite{hri84}; 3 - \cite{kon17}; 4 - \cite{gal16}; 5 - \cite{mun04}; 6 - \cite{gro07}; 7 - \cite{max15}; 8 - \cite{rib99}; 9 - \cite{cla10}; 10 - \cite{and89}; 11 - \cite{and87a};  12 - \cite{hen06}, 13 - \cite{cla08b}; 14 - \cite{bak08}; 15 - \cite{and93};  16 - \cite{slac12}; 17 - \cite{amm06}; 18 - \cite{and87b}; 19 - \cite{pop81};  20 - \cite{tor00}; 21 - \cite{tor09}; 22 - \cite{gra16}; 23 - \cite{vos12}}       
\tablenotetext{a}{Combined, extinction-corrected, out-of-eclipse magnitudes of both components expressed in Johnson photometric system.}  
\tablenotetext{b}{Calculated using the WD model.}    
\tablenotetext{c}{Photometric distances derived from bolometric flux scaling.}                                                 
\end{deluxetable*}
\end{turnpage}

\subsection{Distances}
\subsubsection{Geometric distances}
\label{geom:dist}
The source of parallaxes is almost exclusively
the recent release of Gaia parallaxes DR1 \citep{gaia16} and in
a few cases of close and bright systems where those parallaxes are
unavailable we use parallaxes from the latest reduction of the Hipparcos
data \citep{vLe07}. Distances are calculated through simple inversion
of trigonometric parallaxes. It is known that this procedure for
larger parallax errors ($\gtrsim4\%$) is not unequivocal and must
include some prior on expected space distribution of an object
\citep[e.g.][]{san02,bai15}. Existence of this prior is necessary to
recover a true distribution (distances) from an observed distribution
(parallaxes) in the presence of observational errors. In terms of
absolute luminosity bias it leads to the so called Lutz-Kelker correction
\citep{lut73}. However errors given by the Gaia DR1 are preliminary and
likely overestimated \citep[e.g.][]{cas16} and using them for parallax
corrections would introduce unknown amounts of systematics. For the purpose
of this paper we decided to not apply Lutz-Kelker corrections to the
distances, especially as any such correction would be smaller than quoted
errors. The resulting distances are summarized in Tab.~\ref{tab:fot}.

\subsubsection{Photometric distances}
\label{phot:dist}
We employed the so-called standard method utilizing V-band bolometric corrections to derive
photometric distances, known also as the bolometric flux scaling. We
calculated distance $d$ to the $i$-th component of the system using
equation: 
\begin{equation}
d_{i} ({\rm pc}) = 3.360\cdot10^{-8} R_{i}\, T^2\!\!\!_{i}\; 10^{0.2(BC_{i} + V_{i})}, 
\label{eq:1}
\end{equation}
where index $i=\{1,2\}$, $R$ is the radius of a component in solar radii,
$T$ is its effective temperature in K, $BC$ is a bolometric correction
interpolated from the \cite{flo96} tables for a given temperature and $V$
is the intrinsic magnitude of a component (corrected for extinction). The
distance to a particular system was calculated as the unweighted average distance
of the two components. The purposes of
introducing photometric distances is to check for consistency of the
eclipsing binary model parameters and validation of the Gaia parallaxes 
used in the analysis. The photometric distances are given in 
Tab.~\ref{tab:fot}. 

\subsection{Angular diameters}
\label{sec:ang}
In order to derive surface
brightness -- color relations we need to calculate individual angular
diameters of the stars. Angular diameters are calculated with the formula:
\begin{equation}
\phi ({\rm mas}) = 9.3004 \cdot R ({\rm R_\odot}) / d ({\rm pc}),
\label{eq:2}
\end{equation} 
where $d$ is a distance, $R$ the radius of the star and the conversion factor is equal to $ 2000 R_\odot 
/{\rm 1AU}$ assuming a solar radius $R_\odot=695660$ km \citep{hab08}
and a length of the astronomical unit $ {\rm 1AU} = 149597871$ km \citep{pit09}.

{We emphasize that angular diameter calculated from the
photometric distance is a function of {\it radiative} properties of a
star (mainly its effective temperature) and not its {\it geometric} properties. 
Indeed, if we combine equations
\ref{eq:1} and \ref{eq:2} we derive angular diameter which is
only a function of the effective temperature, the $V$-band bolometric correction
(also temperature dependent) and extinction corrected $V$-band
magnitude. Because of this we do not utilize the photometric distances 
to calculate angular diameters in the present work.

\subsection{Intrinsic magnitudes}
\label{sec:intrin}
In Table~\ref{tab:fot} we summarize all parameters used to derive the intrinsic
photometric indexes of components. The mean galactic extinction curve
with $R_V=3.1$ \citep{fit07} was used to correct the observed magnitudes
for reddening. Next, with the help of our WD models, we calculated
light ratios in the Johnson $BVJHK$ bands and use them to derive intrinsic
magnitudes and colors of each component. The WD code uses an atmospheric
approximation with intensities based on ATLAS9 \citep{kur93} model stellar
atmospheres which are integrated over a given passband to give emerging
flux being expressed as fraction of flux emerging from the black body
of the same temperature. For all the systems in our sample $B$ and $V$
light ratios are tuned to published light ratios based on literature
light curve solutions. However in order to calculate the light ratios in
the infrared $JHK$ bands we need to extrapolate the models as none of
the systems has infrared light curves published or analyzed. This is why
the temperature ratio needs to be well established in order to minimize
systematics due to the extrapolation. Provided the temperature ratio and
absolute temperatures are well known such a procedure does not introduce
significant bias because the relative fluxes from the atmospheric models are
much better constrained than the absolute fluxes. We add also that errors
given on unreddened magnitudes in Table~\ref{tab:fot} do not account
for possible systematic shifts on a level of 1\% due to transformation
of Tycho-2 and 2MASS magnitudes onto the Johnson photometric system.

\subsection{Surface brightness}
\label{sec:defin}
 We follow \cite{hin89}
to define the surface brightness parameter $S$: 
\begin{equation}
 S_i = m_{i,0} + 5 \log{\phi}, \label{equ:sb} 
\end{equation}
where $i$ denotes
a particular band ($B$ or $V$) and $m_{i,0}$ is the intrinsic magnitude
in a given band.  The surface brightness parameter $S$ was then used
to obtain the SBC relations by fitting it with first and fifth degree
polynomials in a form: 
\begin{equation}
S = \sum_{i=0}^{i=1,5} a_i X^i \label{equ:fit} 
\end{equation} 
where $X$ is a given photometric color (see
Sect.~\ref{sec:SBC} for more details). Use of a higher order polynomial is
justified by a strong non-linearity of the SBC relations for the blue-most
colors (stellar spectral types earlier than A0).

\begin{figure}
\mbox{\includegraphics[width=0.49\textwidth]{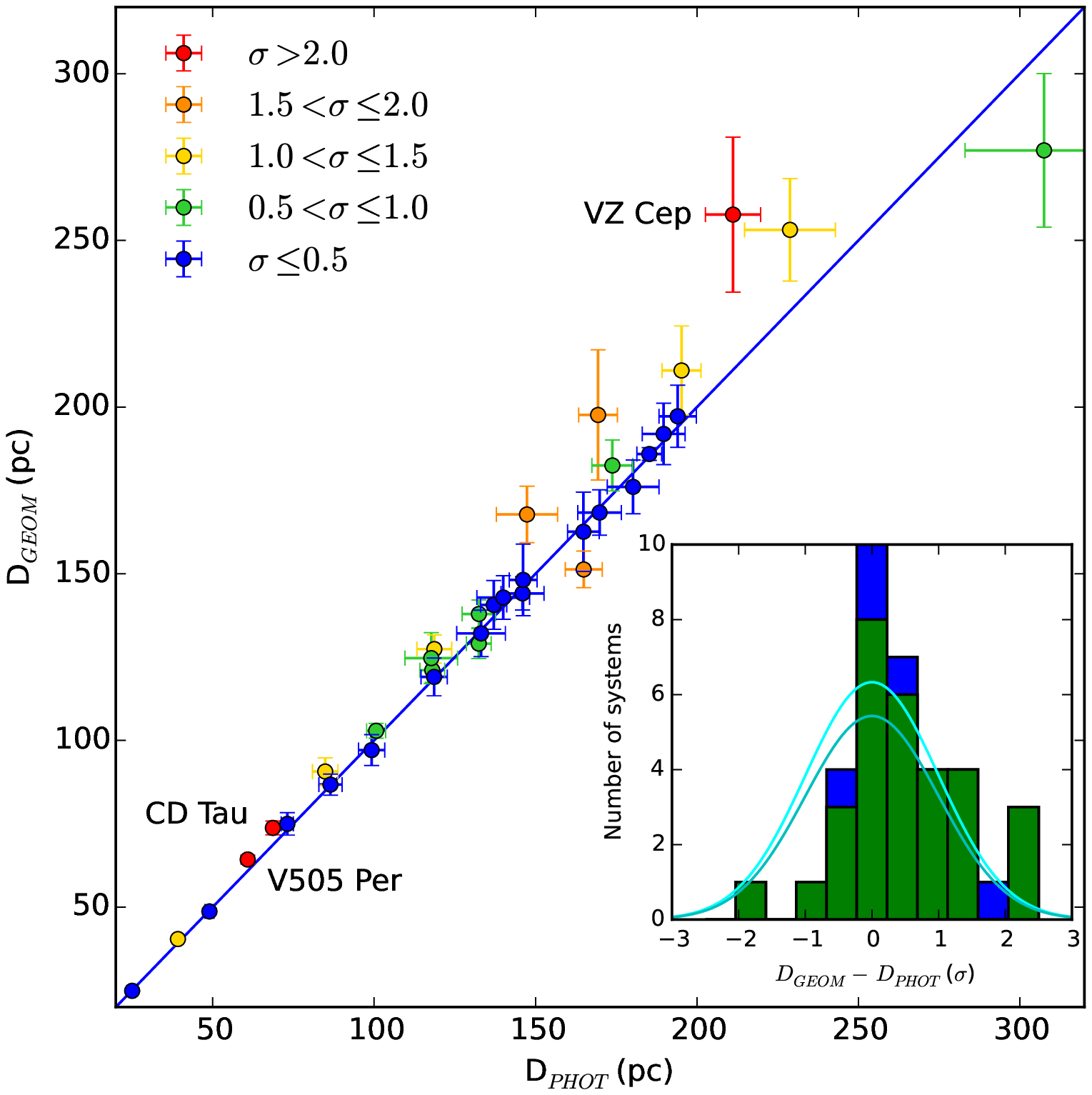}} 
\caption{Comparison of geometric and photometric distances for all systems with deviation from a 1:1 relation expressed as fraction of $\sigma$ and coded with color. Three named systems exhibit offsets larger than 2$\sigma$. Systems with $\sigma\leq0.5$ define the best-fit subsample. {\it Inset}: the expected distribution of deviations from a 1:1 relation for all systems (upper line) and systems with the Gaia parallaxes (lower line) when random errors dominate. The histogram shows the actual distribution of deviations for the Gaia subsample (green) and the entire sample (green+blue).
\label{fig:dist}}
\end{figure}

\begin{figure}
\mbox{\includegraphics[width=0.49\textwidth]{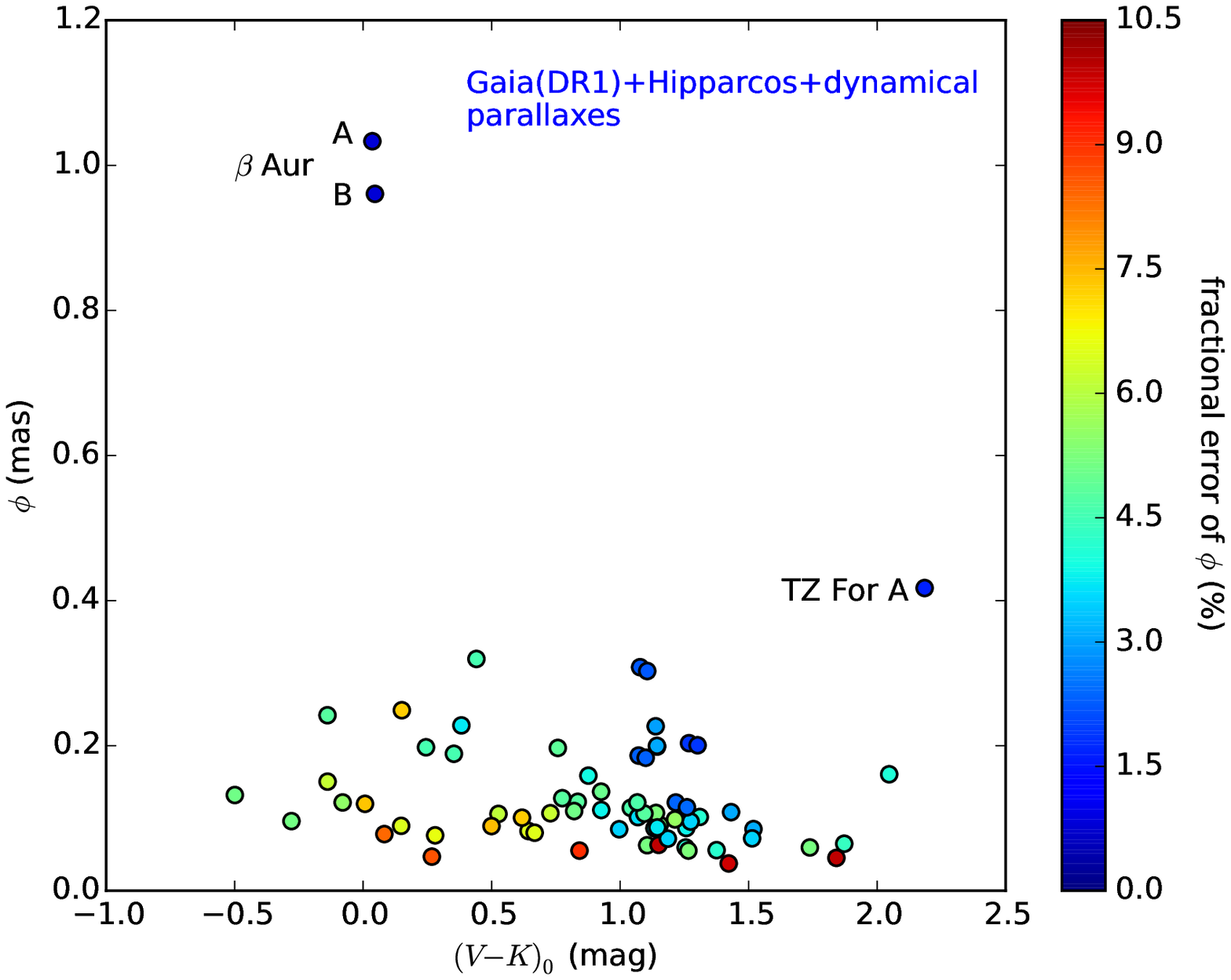}}
\caption{Geometric angular diameters of the eclipsing binary components and their uncertainties calculated from the most recent stellar radii and parallaxes. Three stars that are resolvable by interferometry are named.
\label{fig:angular}}
\end{figure}

\section{Results}\label{results}
\subsection{Distances}
\cite{sta16a,sta16b} presented comprehensive comparisons of geometric
distances from the Hipparcos and Gaia satellites with photometric distances
derived from eclipsing binaries. We underline that we use here a different
method to derive photometric distances, and our sample is significantly
smaller but on other hand more carefully selected. Figure~\ref{fig:dist}
shows the comparison of geometric and photometric distances, where difference
between both distances is expressed in term of the standard deviation
of distances $\sigma$. Inspection of the figure confirms that detached
eclipsing binary stars serve as almost perfect distance indicators, and
the photometric distance is a very good proxy of true geometric distance,
as long as issues with reddening and temperature are properly handled
\citep[e.g. see Section 4 in][]{tor10}. Both distances and their $\sigma$
difference are given in Tab.~\ref{tab:fot}.

The largest deviations from the 1:1 relation between geometric and photometric
distance are for V505~Per (2.2$\sigma$), CD~Tau (2.2$\sigma$) and 
VZ~Cep(2.1$\sigma$). For the complete sample the reduced $\chi^2=0.99$ (34 degrees
of freedom) i.e.~it is fully consistent with statistical uncertainties
dominating the error budget in the distance determination. However if we exclude
the three most deviating systems the reduced $\chi^2=0.87$ (31 degrees of
freedom) suggest that for the majority of systems the errors on the distance
determinations are slightly overestimated. We note that the two strongly
deviating systems (CD~Tau and VZ~Cep) also stand out in comparison of
photometric and Hipparcos parallaxes \citep{sta16a}. A possible explanation
is wrongly estimated temperatures or/and interstellar extinctions in
case of those two systems or the presence of some additional systematics in
trigonometric parallaxes. However more work is needed to figure out the
source of the discrepancy.

The inset in Figure~\ref{fig:dist} shows histogram of deviations in terms of
the $\sigma$ from a sub-sample with Gaia parallaxes (30 systems, green) 
and from the rest of the sample (green+blue). Superimposed are the expected 
distributions of deviations when errors are uncorrelated and dominated by random 
uncertainties. We see a clear excess of systems with small deviations signifying 
that errors on distances are inflated both for the sample and
the Gaia sub-sample. This is in agreement with the conclusion by \cite{cas16}.    
We see also that Gaia distances are on average larger than
photometric distances, thus corroborating findings by \cite{sta16b}.

\subsection{Angular diameters}
\label{sec:angula}
The distances were utilized to calculate geometric angular diameters for all
the sample -- see Tab.~\ref{tab:ang}. Those angular diameters are
direct limb darkened angular diameters and they are complimentary to
angular diameters derived from interferometry \citep[e.g. compilations
by][]{boy14,cha14}. They have an average precision of 4.7\% limited by the
precision of parallax determinations. The precision is better than 2\% for
11 components. Figure~\ref{fig:angular} shows the derived angular diameters
with uncertainties. One can note the clear dependency of uncertainty on
angular size and color $(V-K)_0$, with bluer (hotter) stars having angular
diameters more poorly determined. From all the sample only one star, the
cooler component of TZ~For, had its angular diameter measured directly
with interferometry \citep{gal16}, however with much lower precision,
and two components of $\beta$ Aur were barely resolvable \citep{hum95}. 

\begin{deluxetable*}{@{}lccc@{$\,\pm\,$}cc@{$\,\pm\,$}cc@{$\,\pm\,$}cc@{$\,\pm\,$}cc@{$\,\pm\,$}cc@{$\,\pm$}c@{}}
\tabletypesize{\scriptsize}
\tablecaption{Metallicities from literature and derived quantities: masses, gravities and geometric angular diameters of all eclipsing binary components.  \label{tab:ang}}
\tablewidth{0pt}
\tablehead{\colhead{ID} & \colhead{[Fe$/$H]} & Ref & \multicolumn{4}{c}{Mass} & \multicolumn{4}{c}{Gravity} & \multicolumn{4}{c}{Angular Diameter} \\
\colhead{} & & &\multicolumn{2}{c}{$M_1 \pm \sigma$}&\multicolumn{2}{c}{$M_2 \pm \sigma$}&  \multicolumn{2}{c}{$\log{g_1} \pm \sigma$}&\multicolumn{2}{c}{$\log{g_2} \pm \sigma$} & \multicolumn{2}{c}{$\theta_1 \pm \sigma$} & \multicolumn{2}{c}{$\theta_2 \pm \sigma$} \\
\colhead{} & & &\multicolumn{2}{c}{$M_\sun$} & \multicolumn{2}{c}{$M_\sun$} & \multicolumn{2}{c}{dex} &\multicolumn{2}{c}{dex} & \multicolumn{2}{c}{mas} & \multicolumn{2}{c}{mas}}
\startdata
YZ Cas  & 0.10 & 1 & 2.263 & 0.012 &  1.325 & 0.007 & 3.988 & 0.004 &  4.311 & 0.005& 0.242 & 0.012 &  0.128 & 0.006 \\
AI Phe  & -0.14 & 2& 1.193 & 0.004 &  1.242 & 0.004 & 3.985 & 0.008 &  3.605 & 0.004& 0.102 & 0.004 &  0.161 & 0.007  \\
V505 Per  & -0.12 & 3& 1.272 & 0.003 &  1.254 & 0.003 & 4.323 & 0.009 &  4.331 & 0.010& 0.186 & 0.004 &  0.183 & 0.004  \\
AL Ari  & -0.00 &4 & 1.170 & 0.006 &  0.916 & 0.004 & 4.230 & 0.004 &  4.489 & 0.004& 0.091 & 0.005 &  0.060 & 0.003 \\
V570 Per  & 0.02 &5 & 1.452 & 0.009 &  1.352 & 0.009 & 4.234 & 0.017 &  4.283 & 0.012& 0.111 & 0.004 &  0.101 & 0.004 \\
\\
TZ For  & 0.02 & 6& 1.957 & 0.002 &  2.056 & 0.002 & 3.532 & 0.018 &  2.909 & 0.011& 0.199 & 0.005 &  0.417 & 0.007  \\
V1229 Tau  & 0.06 &7 & 2.203 & 0.013 &  1.549 & 0.010 & 4.306 & 0.014 &  4.273 & 0.026& 0.122 & 0.007 &  0.106 & 0.006  \\
V1094 Tau  & -0.09 & 8& 1.096 & 0.004 &  1.012 & 0.003 & 4.179 & 0.004 &  4.355 & 0.006& 0.108 & 0.003 &  0.085 & 0.003  \\
CD Tau  & 0.08 & 9& 1.441 & 0.016 &  1.367 & 0.016 & 4.087 & 0.010 &  4.174 & 0.012& 0.227 & 0.007 &  0.200 & 0.006  \\
EW Ori  & 0.05 & 10& 1.177 & 0.009 &  1.130 & 0.008 & 4.373 & 0.005 &  4.409 & 0.005& 0.060 & 0.003 &  0.056 & 0.002  \\
\\
UX Men  & 0.04 &11 & 1.229 & 0.006 &  1.192 & 0.007 & 4.270 & 0.009 &  4.305 & 0.009& 0.122 & 0.003 &  0.115 & 0.003  \\
TZ Men  & -- & & 2.482 & 0.025 &  1.500 & 0.010 & 4.224 & 0.010 &  4.302 & 0.010& 0.150 & 0.009 &  0.107 & 0.007  \\
beta Aur  & 0.15 & 12& 2.365 & 0.006 &  2.303 & 0.006 & 3.929 & 0.005 &  3.981 & 0.005& 1.033 & 0.008 &  0.961 & 0.008  \\
RR Lyn  & -0.24 & 13& 1.922 & 0.026 &  1.504 & 0.041 & 3.900 & 0.009 &  4.214 & 0.020& 0.320 & 0.015 &  0.197 & 0.010  \\
WW Aur  & -- & & 1.964 & 0.010 &  1.814 & 0.008 & 4.161 & 0.005 &  4.167 & 0.006& 0.198 & 0.009 &  0.189 & 0.009  \\
\\
HD71636  & -0.05 &14 & 1.512 & 0.007 &  1.285 & 0.006 & 4.226 & 0.005 &  4.279 & 0.005& 0.123 & 0.006 &  0.106 & 0.005  \\
VZ Hya  & -0.20 & 15& 1.271 & 0.009 &  1.146 & 0.006 & 4.305 & 0.005 &  4.404 & 0.006& 0.085 & 0.003 &  0.072 & 0.003  \\
KX Cnc  & 0.07 &16 & 1.142 & 0.003 &  1.132 & 0.003 & 4.441 & 0.002 &  4.450 & 0.003& 0.203 & 0.004 &  0.201 & 0.004  \\
PT Vel  & -- & & 2.199 & 0.016 &  1.626 & 0.009 & 4.138 & 0.009 &  4.264 & 0.011& 0.120 & 0.009 &  0.089 & 0.007  \\
KW Hya  & -- & & 1.973 & 0.036 &  1.485 & 0.017 & 4.078 & 0.010 &  4.269 & 0.014& 0.228 & 0.008 &  0.159 & 0.006  \\
\\
RZ Cha  & -0.02 &17 & 1.505 & 0.027 &  1.513 & 0.021 & 3.946 & 0.011 &  3.893 & 0.010& 0.114 & 0.005 &  0.122 & 0.006 \\
FM Leo  & -- & & 1.318 & 0.011 &  1.287 & 0.010 & 4.124 & 0.023 &  4.189 & 0.028& 0.107 & 0.006 &  0.098 & 0.006  \\
GG Lup  & -0.10 & 18& 4.079 & 0.039 &  2.508 & 0.022 & 4.295 & 0.009 &  4.360 & 0.009& 0.132 & 0.007 &  0.096 & 0.005  \\
V335 Ser  & -- & & 2.147 & 0.014 &  1.905 & 0.008 & 4.155 & 0.011 &  4.240 & 0.017& 0.089 & 0.006 &  0.076 & 0.005 \\
WZ Oph  & -0.27 & 15& 1.227 & 0.007 &  1.220 & 0.006 & 4.233 & 0.008 &  4.220 & 0.008& 0.086 & 0.003 &  0.087 & 0.003  \\
\\
FL Lyr  & -0.30 &19 & 1.218 & 0.016 &  0.958 & 0.012 & 4.307 & 0.021 &  4.453 & 0.028& 0.087 & 0.003 &  0.065 & 0.003  \\
UZ Dra  & -- & & 1.306 & 0.012 &  1.203 & 0.011 & 4.330 & 0.017 &  4.402 & 0.019& 0.063 & 0.003 &  0.055 & 0.003  \\
V4089 Sgr  & -- & & 2.584 & 0.012 &  1.607 & 0.008 & 3.654 & 0.005 &  4.233 & 0.004& 0.249 & 0.018 &  0.101 & 0.007  \\
V1143 Cyg  & 0.08 & 20& 1.356 & 0.003 &  1.328 & 0.002 & 4.317 & 0.015 &  4.323 & 0.015& 0.308 & 0.007 &  0.303 & 0.007  \\
MY Cyg  & -- & & 1.806 & 0.025 &  1.782 & 0.030 & 3.994 & 0.020 &  4.013 & 0.021& 0.082 & 0.005 &  0.080 & 0.005  \\
\\
EI Cep  & -0.04 &21 & 1.772 & 0.006 &  1.680 & 0.006 & 3.762 & 0.014 &  3.928 & 0.016& 0.137 & 0.007 &  0.110 & 0.006  \\
VZ Cep  & 0.06 & 22& 1.402 & 0.015 &  1.108 & 0.008 & 4.213 & 0.008 &  4.446 & 0.033& 0.055 & 0.005 &  0.038 & 0.004  \\
LL Aqr  & 0.02 & 23& 1.195 & 0.001 &  1.034 & 0.001 & 4.272 & 0.004 &  4.451 & 0.004& 0.095 & 0.003 &  0.072 & 0.003  \\
EF Aqr  & 0.00 & 24& 1.243 & 0.006 &  0.946 & 0.003 & 4.276 & 0.007 &  4.447 & 0.007& 0.063 & 0.006 &  0.045 & 0.004  \\
V821 Cas  & -- & & 2.088 & 0.064 &  1.655 & 0.050 & 4.024 & 0.017 &  4.364 & 0.024& 0.078 & 0.007 &  0.047 & 0.004 
\enddata
\tablecomments{References to metallicities:  1 - \cite{pav14}; 2 - \cite{and88}; 3 - \cite{tom08a};  4 - \cite{kon17}; 5 - \cite{tom08b}; 6 - \cite{gal16};  7 - \cite{gro07}; 8 - \cite{max15}; 9 - \cite{rib99}; 10 - \cite{cla10}; 11 - \cite{and89}; 12 - \cite{sou07}; 13 - \cite{kha01}; 14 - \cite{hol09}; 15 - \cite{cla08b}; 16 - \cite{sow12}; 17 - \cite{jor75}; 18 - \cite{and93}; 19 - \cite{gui09}; 20 -  \cite{and87b};  21 - \cite{tor00}; 22 - \cite{tor09}; 23 - \cite{gra16}; 24 - \cite{vos12}} 
\end{deluxetable*}

\begin{deluxetable*}{ccccccccccc}
\tabletypesize{\scriptsize}
\tablecaption{Coefficients of polynomial fits to the Surface Brightness parameter $S$ in $B$- and $V$-bands.  \label{tab:calib}}
\tablewidth{0pt}
\tablehead{
\colhead{Band} & \colhead{Color}  & \colhead{$N\tablenotemark{a}$} & \colhead{Color Range} & \colhead{$a_0$} &\colhead{$a_1$} & \colhead{$a_2$}&\colhead{$a_3$} &\colhead{$a_4$} & \colhead{$a_5$} & \colhead{$\sigma$} \\
\colhead{} & \colhead{Index}  & \colhead{} & \colhead{(mag)} & \colhead{} &\colhead{} & \colhead{}&\colhead{} &\colhead{} & \colhead{} & \colhead{$\%$}  
}
\startdata
\multicolumn{11}{c}{Linear fits (best-fit subsample)}\\
$B$&$(B\!-\!K)$&28&[$-$0.12:3.15]&2.640(18)&1.252(11)&-&-&-&-&2.5\\
$V$&$(B\!-\!K)$&28&[$-$0.12:3.15]&2.625(15)&0.959(9)&-&-&-&-&2.2 \\
$V$&$(V\!-\!K)$&28&[$-$0.10:2.15]&2.644(19)&1.358(17)&-&-&-&-&2.7\\
\multicolumn{11}{c}{Fifth-order polynomial fits (entire sample)}\\
$B$&$(B\!-\!K)$&70&[$-$0.7:3.15]&2.594(31)&1.423(88)&$-$0.592(164)&0.612(200)&$-$0.239(93)&0.031(14)&5.1\\
$V$&$(B\!-\!K)$&70&[$-$0.7:3.15]&2.579(27)&1.134(85)&$-$0.598(155)&0.623(187)&$-$0.245(87)&0.032(13)&5.0\\
$V$&$(V\!-\!K)$&70&[$-$0.5:2.15]&2.606(33)&1.526(134)&$-$0.989(317)&1.498(574)&$-$0.835(395)&0.156(88)&5.2
\enddata
\tablenotetext{a}{Number of stars used in the fit}
\tablecomments{Notes. The $S$ parameter is defined by the Equation~\ref{equ:sb}. Colors are in the Johnson photometric system. The limb darkened stellar angular diameter is expressed in milliseconds of arc and follows from the equation: $\log{\theta_{\rm LD}} =  0.2*(a_0 - m +a_1*X+ ... + a_5*X^5)$, where $m$ is the observed extinction-free magnitude of a star in the $B$ or $V$ band and $X$ is an extinction-free color. The last column gives the precision in predicting the angular diameter of stars in the given color range.}
\end{deluxetable*}

\subsection{SBC relations}
\label{sec:SBC}
Figure~\ref{fig:all} shows the relation between the V-band surface brightness
$S_V$ and color $(V-K)_0$ against some interferometric SBC relations
\citep{cha14,boy14,diB05,ker04}. Left and right panels correspond to $S_V$
derived from the complete sample and the best-fit subsample, respectively. The
best-fit systems were defined as those having their geometrical and
photometric distances in agreement to better than $0.5\sigma$ -- see
Tab.~\ref{tab:fot} and Fig.~\ref{fig:dist}. The $V$-band surface brightnesses
derived from trigonometric parallaxes fits well on the \cite{cha14}
calibration with a spread of  $\sim$0.1 mag corresponding to 5\% uncertainty
in angular diameter, dominated by distance errors. The agreement with the
interferometric relation is satisfactory, i.e.~both methods of
measuring angular diameters, direct from interferometry and semi-direct
from eclipsing binary stars, show good consistency. The agreement is
even better if we use the best-fit subsample.

In order to quantify the SBC relation we derived it directly. We
fitted Eq.~\ref{equ:fit} to the $S_V$ (see Sect.~\ref{sec:defin})
using Orthogonal Distance Regression \citep{bog90} which accounts for the errors on  
the independent variable, in our case: color
$(V\!-\!K)_0$. We fitted a fifth-order polynomial to all the data and a first
order polynomial to the data from the best-fit systems.  The results of
the fitting are presented in Figure~\ref{fig:grav} and coefficients of
the derived relations are given in Tab.~\ref{tab:calib}. The precision
of the SBC relation based on all systems is rather low ($\sim 5\%$)
with the distance errors fully dominating the error budget. However, the use
of systems having the best consistency of their geometric and photometric
distances results in a remarkable improvement of the precision of the
derived SBC relation by a factor of 2. The internal precision of the linear
relation in predicting angular diameters of A-, F- and G-type stars
is in fact comparable to or even better than published interferometric
relations up-to-now, e.g.: \cite{boy14} -- 4.6\%, \cite{cha14} --
3.7\%, \cite{ker04} -- 2.8\%\footnote{The precision of $S_V$-$(V\!-\!K)$
calibration by \cite{ker04} is reported to be $1\%$. We recalculated the unweighted
root-mean-square from data given in \cite{ker04}. We obtained a relative
precision of $\sim 3\%$ which we quote the above.} and \cite{diB05} -- 2.1\%. 
The linear SBC relation we derived is almost indistinguishable from the relation by \cite{boy14},
it compares well with the relation by \cite{cha14}, especially for the
bluest colors, and also with \cite{diB05} for the reddest colors 
$(V\!-\!K)_0>1.0$. This is an important argument in favor of the 
eclipsing binary method as a fully independent way to derive the SBC calibration.   

One of the advantages of using eclipsing binary stars comes from the
very precise surface gravities derived for the individual components. 
This allows, in principle, to determine how a SBC relation might depend on surface gravity
(Fig.~\ref{fig:grav}). We see some hints of this dependence where higher
surface gravities result in higher surface brightness but the spread is
still large and it is premature to draw a conclusion here.

The broadband SBC relations calibrated onto a wide range of colors do
not show any statistically significant metallicity dependence with an
exception of the bluest colors \citep[e.g. $(B\!-\!V)$, see][]{boy14}. We
compiled  the metallicity determinations for our sample from the literature
(Tab.~\ref{tab:ang}) in order to check the possible dependence. As 
expected, no clear metallicity dependence is visible for the $V\!-\!K$ color --
see Figure~\ref{fig:met}, although the scatter may hide it.

For the SBC relation to be useful it should have small intrinsic scatter
and be only weakly dependent on reddening. The SBC relation for 
the $V$ band and $(V\!-\!K)$ has great potential in this respect. 
This relation is commonly
used to predict angular diameters and to determine distances, e.g.~to the
Magellanic Clouds with accuracy of 2-3\% \citep{pie13,gra14}. However, for
early-type stars (O or B) the relation becomes non-negligibly inclined
to the line of reddening and shows significantly larger scatter than
for stars with spectral types later than A5 \citep[e.g.][]{cha14}. This reduces
its potential for predicting angular diameters of early type stars. 
\cite{ker04} reported that the SBC relations based on colors
with a larger wavelength difference show smaller scatter, i.e. the colors
$(B\!-\!K)$ and $(V\!-\!L)$. However their relations were constrained
to intermediate- and late-type stars.

\begin{figure*}
\mbox{\includegraphics[width=0.49\textwidth]{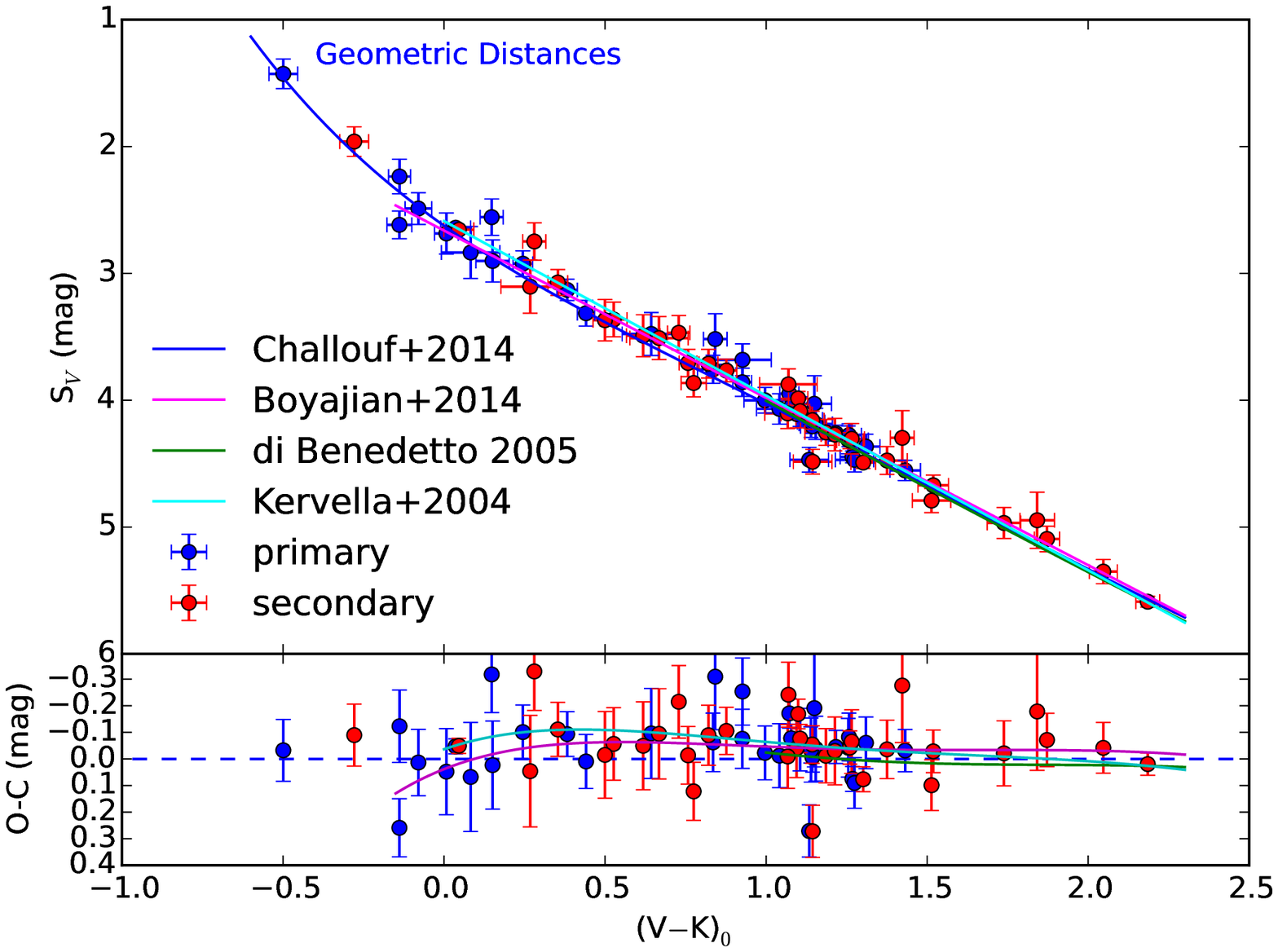}} 
\mbox{\includegraphics[width=0.49\textwidth]{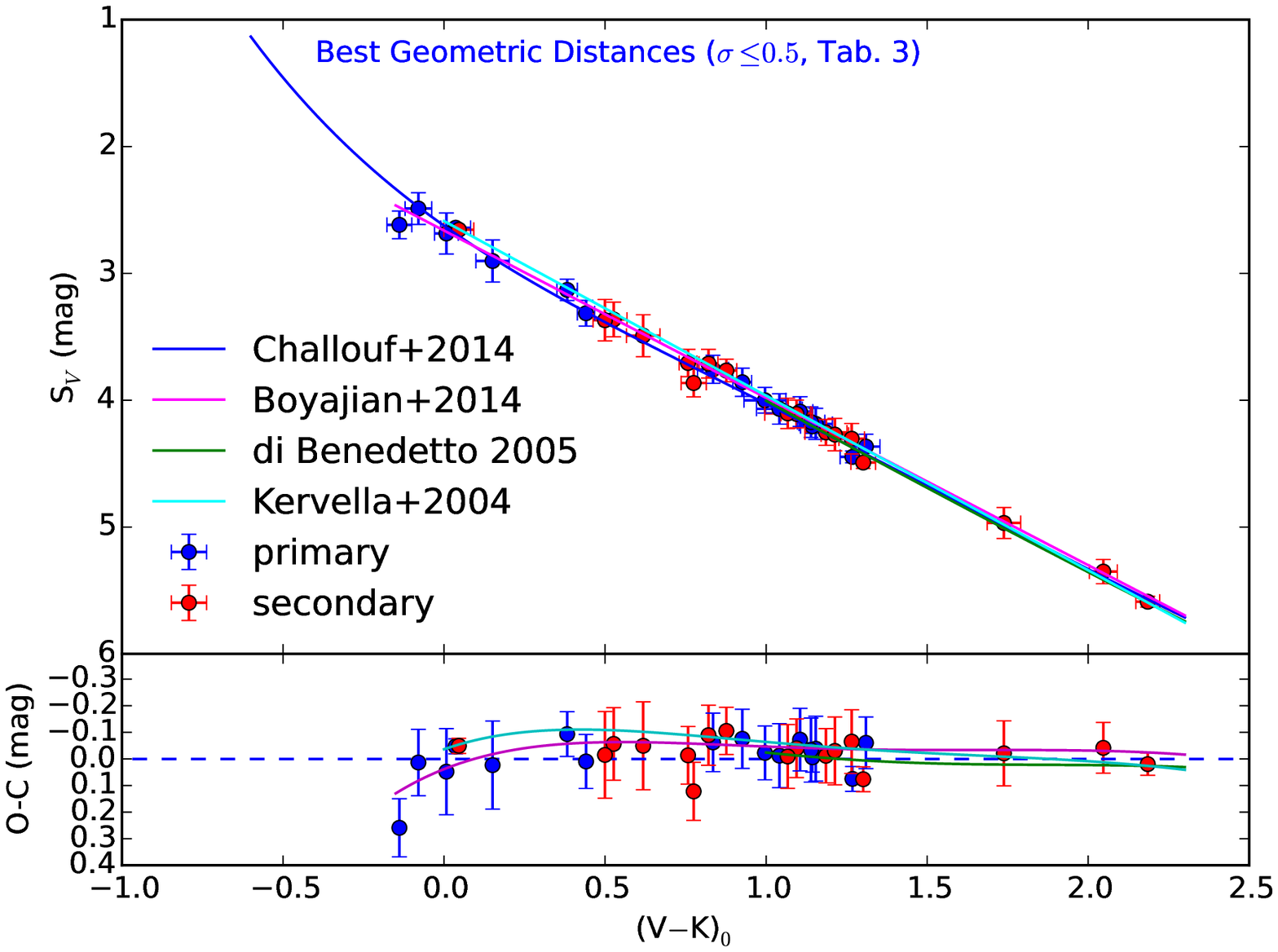}} 
\caption{$V$-band surface brightness vs Johnson color $V\!-\!K$ relation. {\it Left panel}: for all stars based on their geometric distances. {\it Right panel}: for 14 systems with best agreement between the geometric and the photometric distances. Continuous lines correspond to several published interferometric SBC relations. Lower panels show O$\!-\!$C residuals calculated with respect to SBC relation by \cite{cha14} - dashed line. 
\label{fig:all}}

\end{figure*}
\begin{figure*}
\mbox{\includegraphics[width=0.49\textwidth]{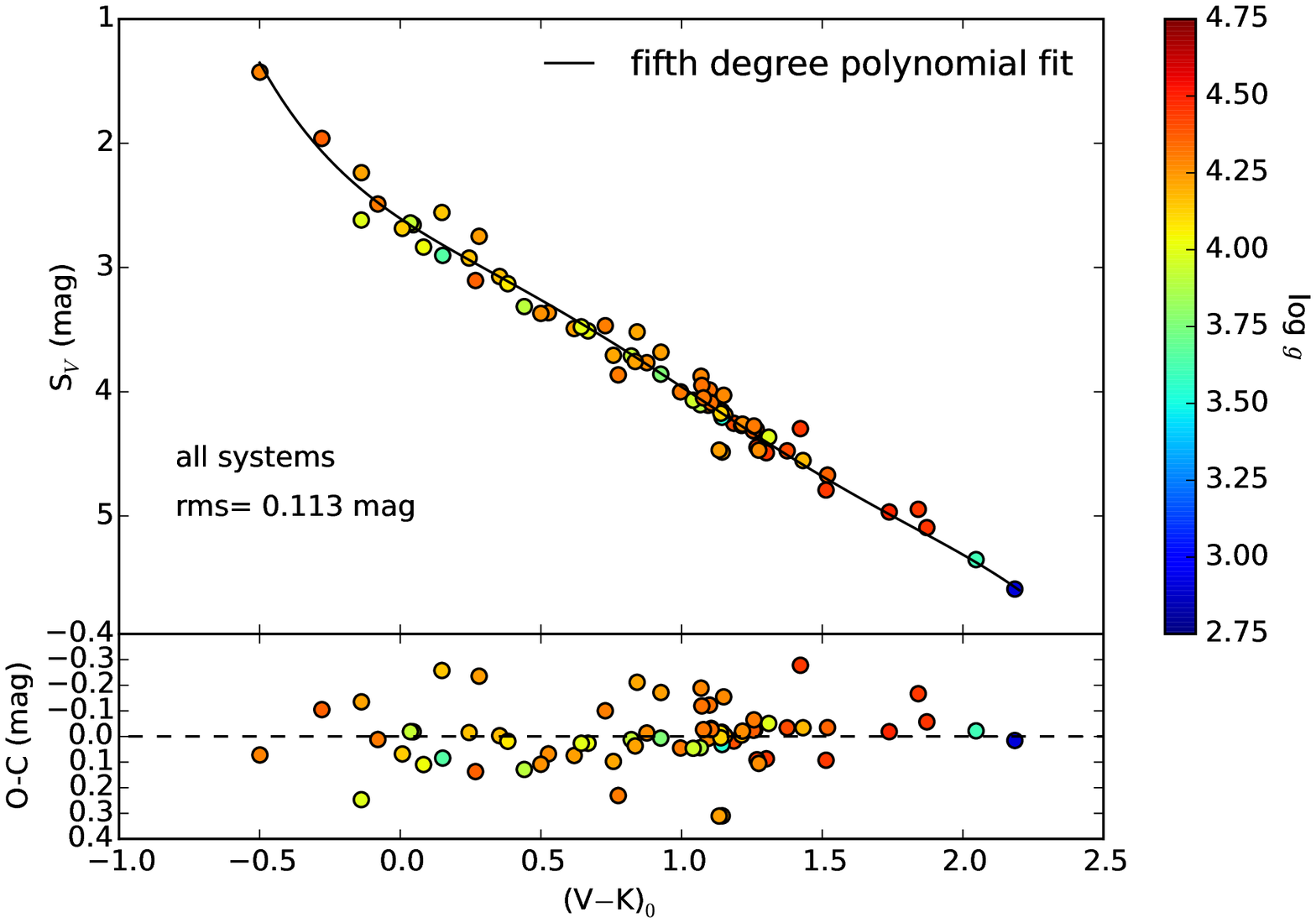}} 
\mbox{\includegraphics[width=0.49\textwidth]{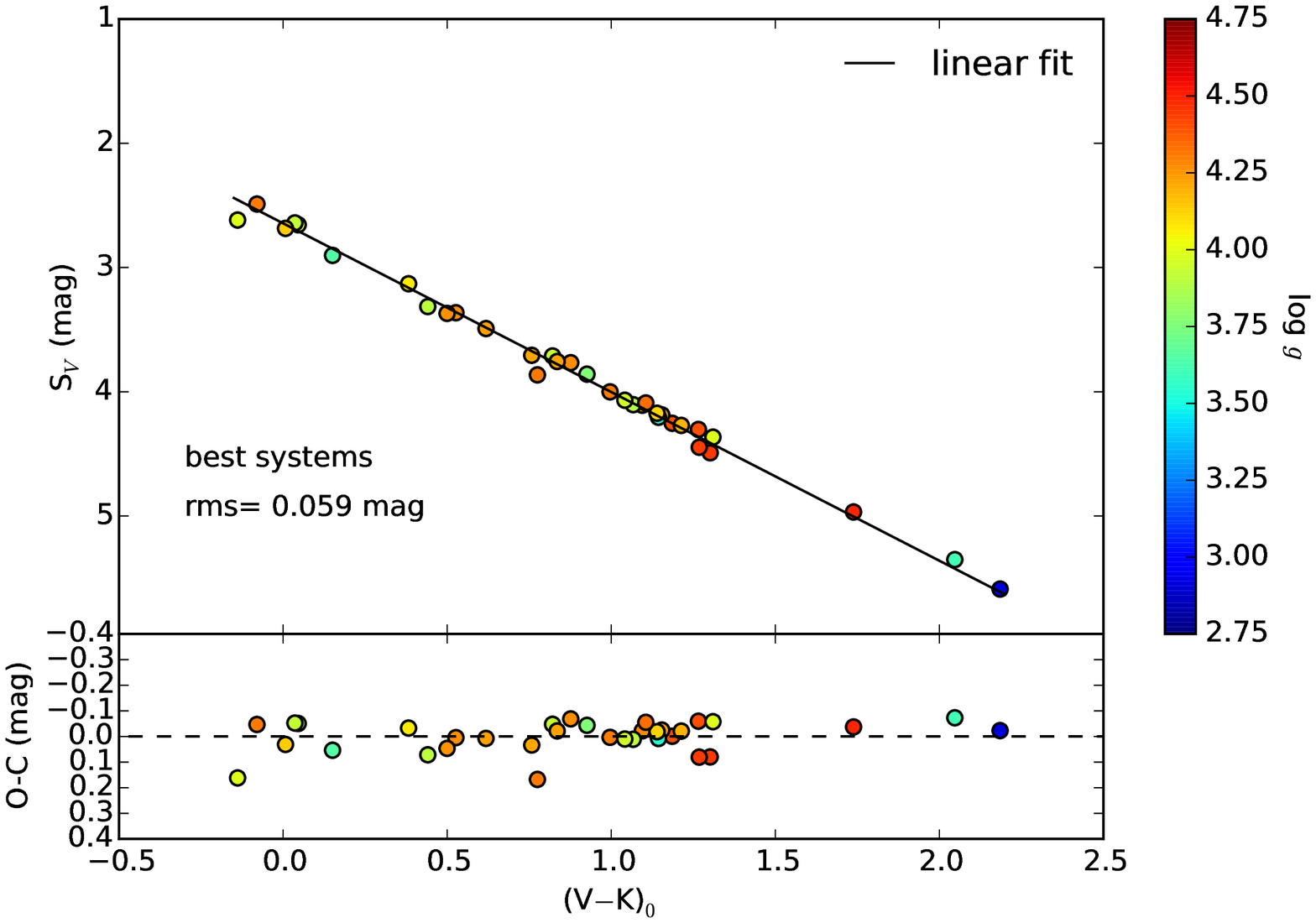}} 
\caption{$V$-band surface brightness vs Johnson color $V\!-\!K$ relation with the addition of the surface gravity color scale (right vertical axis). {\it Left panel}: all the systems. {\it Right panel}: the systems with the best-fit distances.    
\label{fig:grav}}
\end{figure*}

\begin{figure}
\mbox{\includegraphics[width=0.49\textwidth]{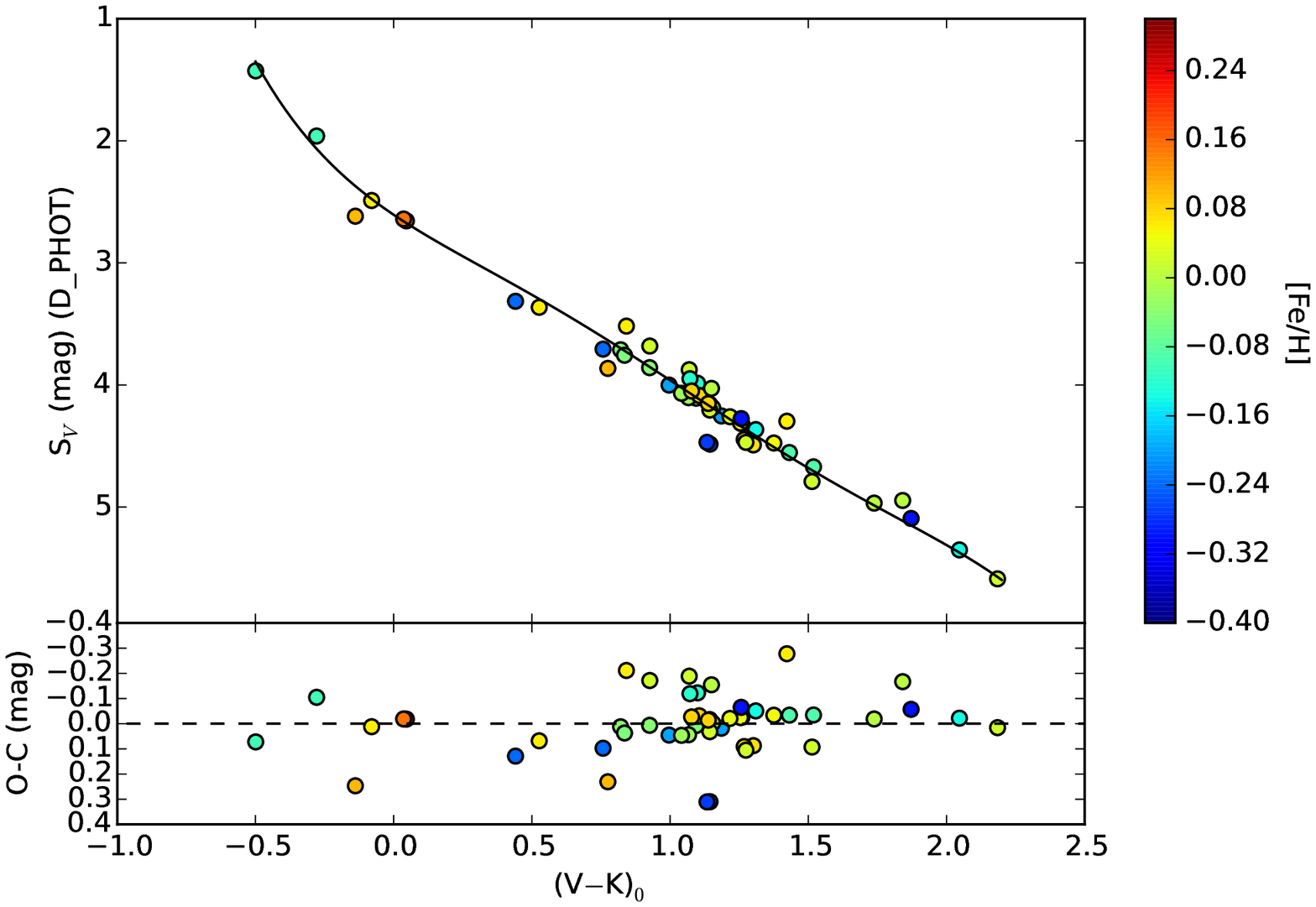}} 
\caption{$V$-band surface brightness vs Johnson color $V\!-\!K$ for systems with determined metallicity. The continous line represents a fifth-order polynomial fit to the entire sample.
\label{fig:met}}
\end{figure}

We decided to search for similar relations using our sample. We combined
two surface brightness parameters ($B$- and $V$-band) with the six colors
$(V\!-\!J)$, $(V\!-\!H)$, $(V\!-\!K)$, $(B\!-\!J)$, $(B\!-\!H)$
and $(B\!-\!K)$. We fit the surface brightness parameters for
the best-fit and all systems using first- and fifth-order polynomials,
respectively. Figure~\ref{fig:BK} shows the two derived promising SBC
relations based on $(B\!-\!K)$ color and with the rms minimized. The
appropriate polynomial coefficients and the precision of angular diameter
prediction are reported in Tab.~\ref{tab:calib}. Both relations give
precisions in the predicted angular diameters of 5\% for the entire sample, 2-3\%
for the best-fit subsample, and have the smallest inclination of the relations
to the reddening line for bluest colors. We note here that the real precision is lower 
because of the global systematic uncertainty of Gaia DR1 parallaxes. 
We estimated that the an upper limit of the systematics is about 3\% for our systems.
 
\begin{figure*}
\mbox{\includegraphics[width=0.49\textwidth]{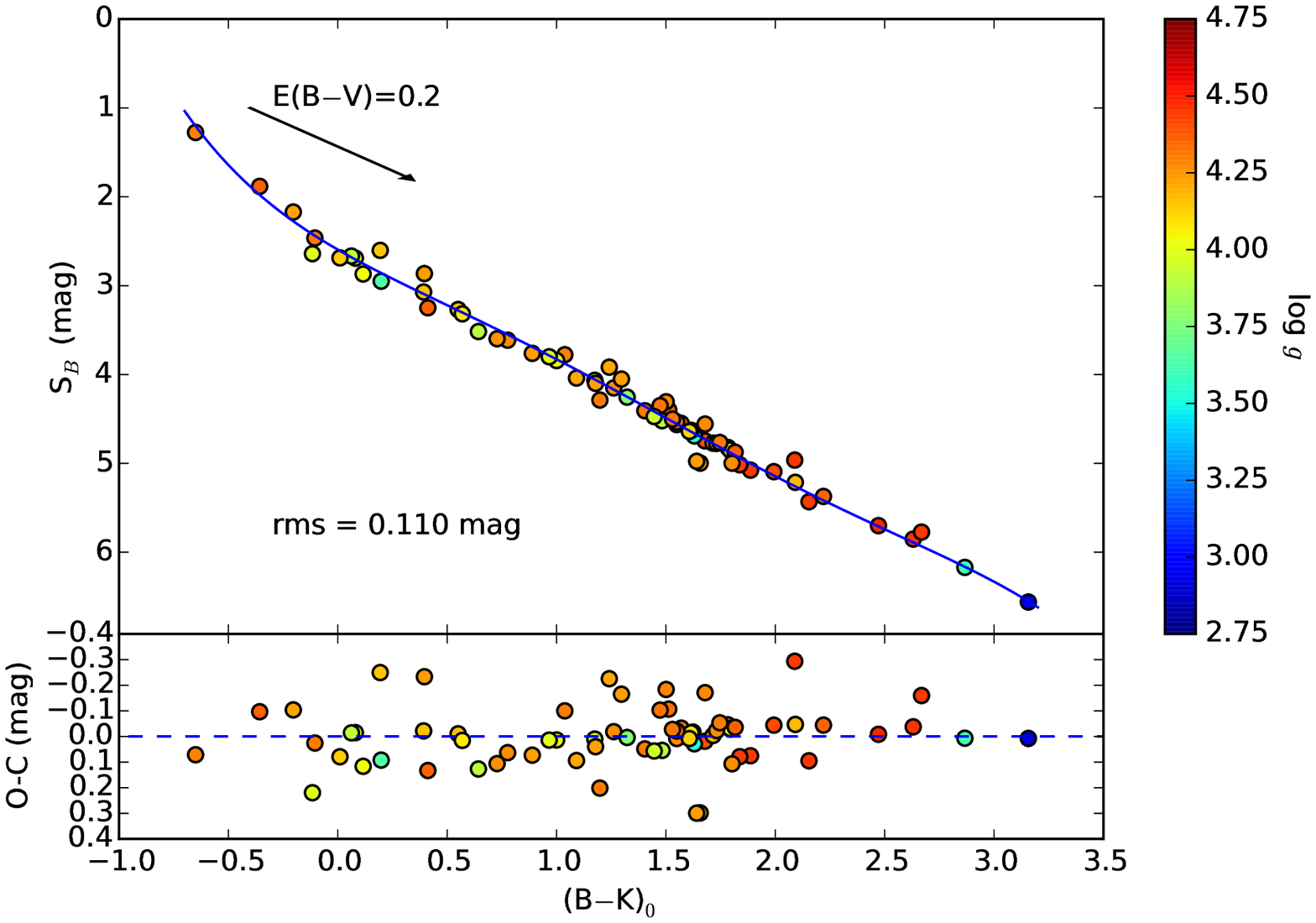}} 
\mbox{\includegraphics[width=0.49\textwidth]{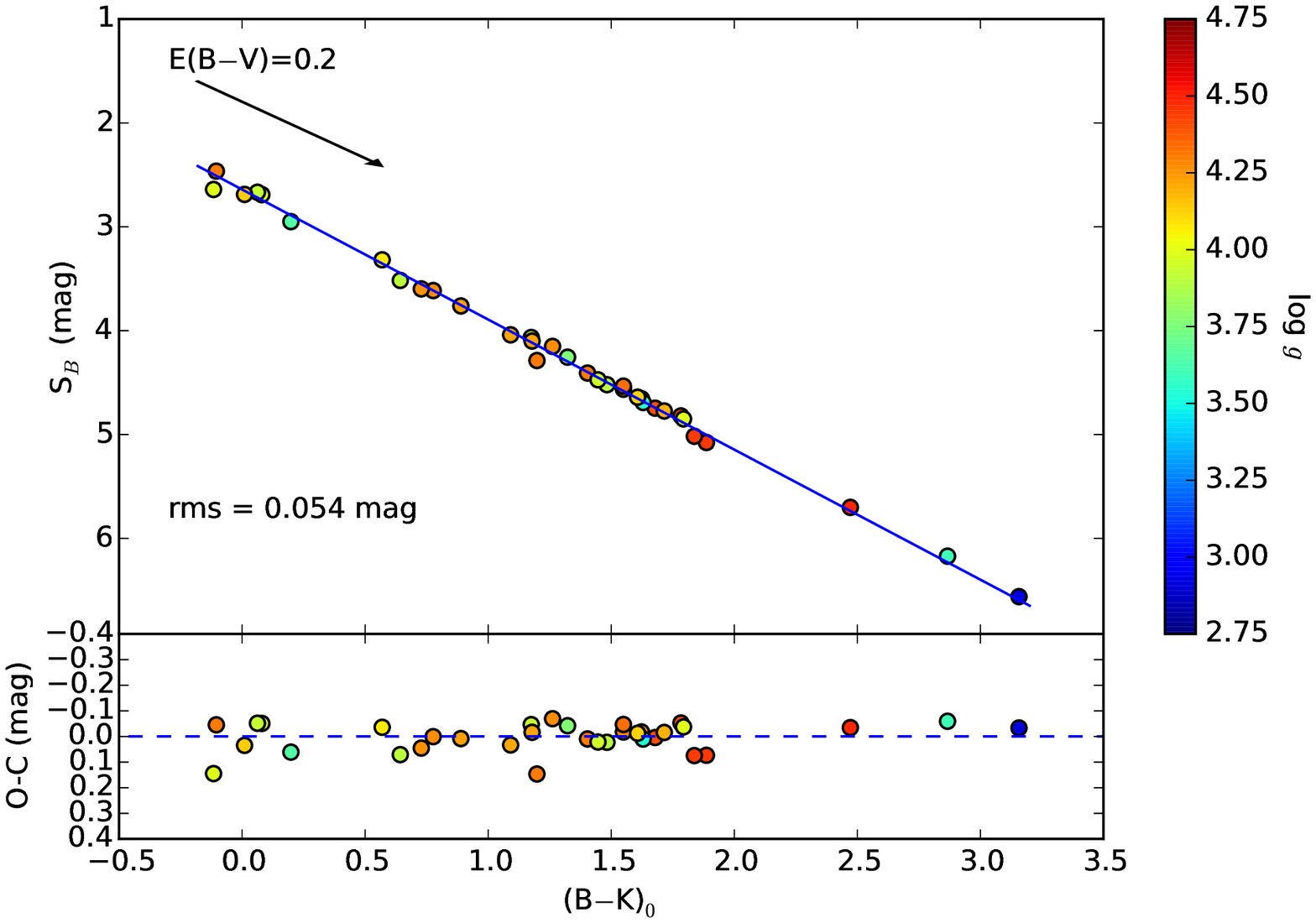}} \\
\mbox{\includegraphics[width=0.49\textwidth]{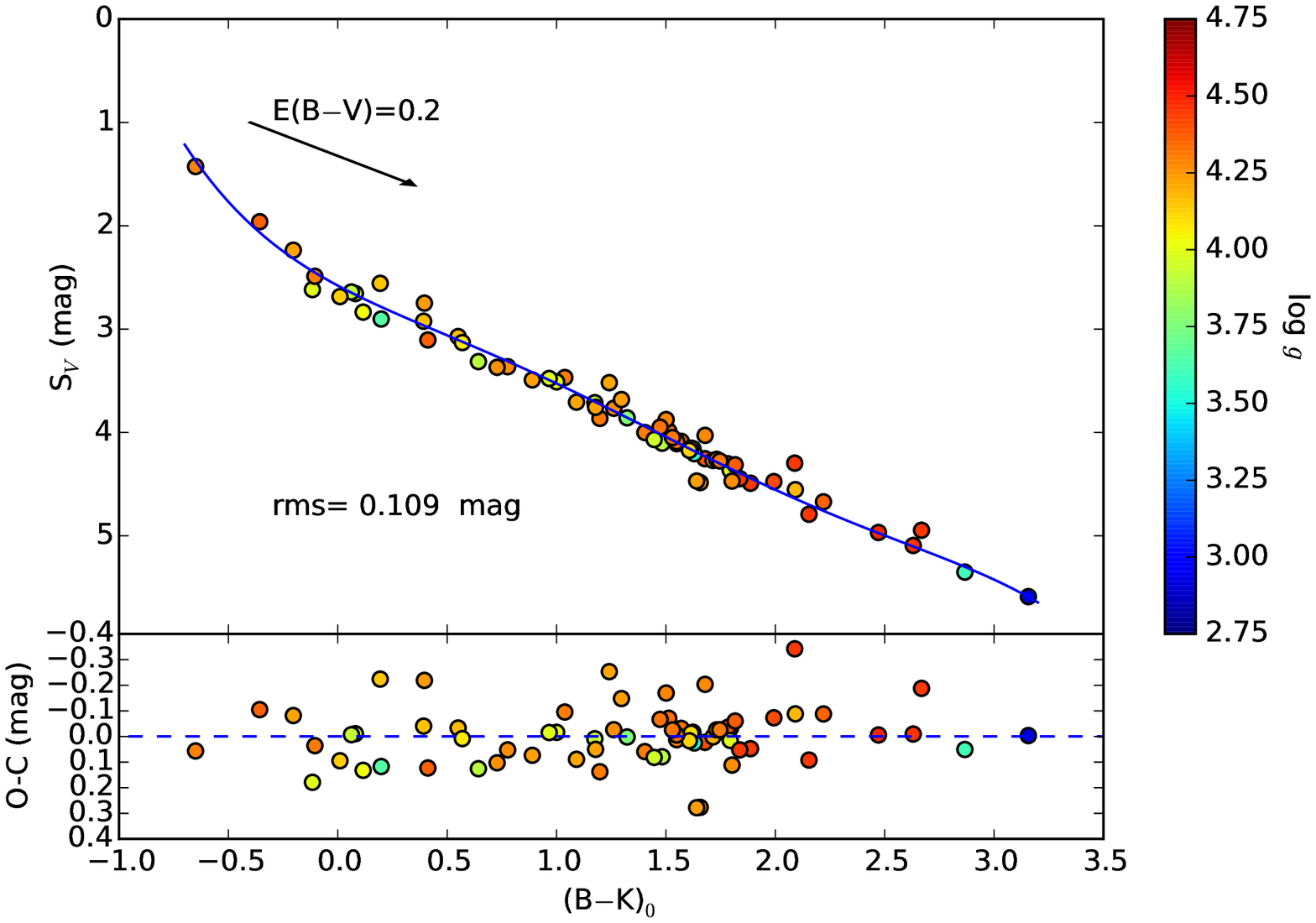}} 
\mbox{\includegraphics[width=0.49\textwidth]{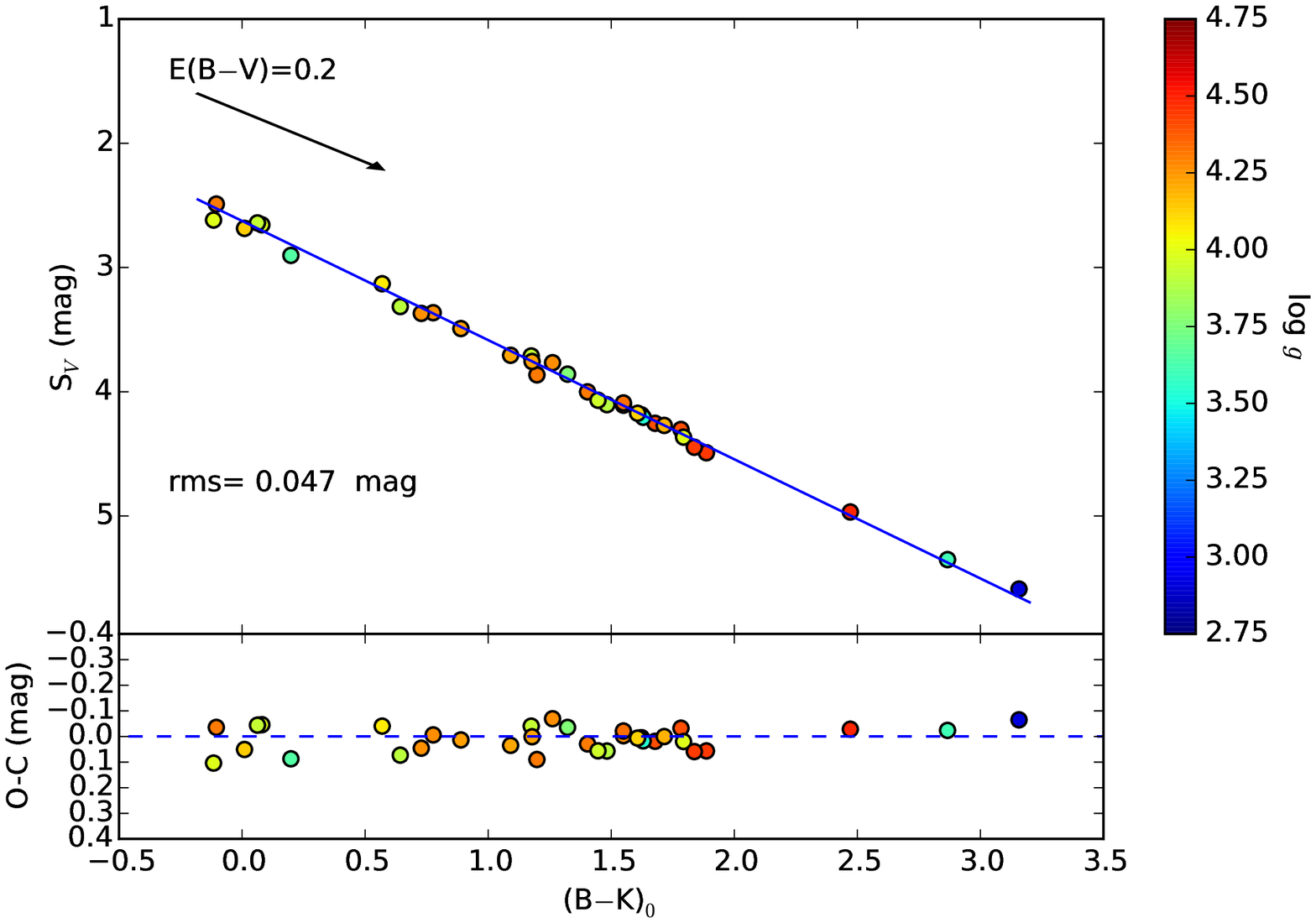}} \\
\caption{Surface brightness vs Johnson color $B\!-\!K$ relations. {\it Upper panel}:
calibrated for the $B$ band. {\it Lower panel}: calibrated for the $V$ band. 
The continuous line
in the left panels shows fifth-order polynomial fits to all stars in
the sample, and in the right panels the line shows linear fits to the best-fit
subsample. The reddening vector is denoted as an arrow. The root mean square of
the relations is given. The surface gravity in $cgs$ units is color coded. 
\label{fig:BK}}
\end{figure*}

\section{Discussion}
\label{discu}
The main purpose of the paper is to
show that the inverse eclipsing binary method allows for independent
and precise calibration of the SBC relations. Results presented in
Section~\ref{sec:SBC} fully corroborate this premise. Still, the precision
of the derived relations is not significantly better than those derived from
interferometric measurements of stellar angular diameters. In this section
we are going to quantify the necessary steps in order to reach sub-percent
precision in predicting angular diameters.

\subsection{Uniform analysis}
We compiled in this work data from numerous
papers published by many different groups of researchers. Each group
uses different quality photometric and spectroscopic data, different
methodology to derive radial velocities, analysis of light and radial
velocity curves (separated, simultaneous, single light curve, multi-band
light curves), different ways of deriving effective temperatures
(color-temperature calibrations, atmospheric model analysis) and finally
different sets of astrophysical numerical constants. During this work we
made some effort to homogenize existing data on eclipsing binary stars,
but it was constrained to the effective temperatures, their ratio and the
radial velocity semi-amplitudes. 

However, to pin-down systematics
a full homogenous re-analysis of each system would be needed using the same 
methodology and software, and also similar quality observables. 
That would result in better evaluation of relative
precision of each data sets and it would augmented the internal precision of
the physical parameters of the whole sample. Significant help in this
respect can be expected from using new, high precision numerical codes like {\it ellc} \citep{max16} or Phoebe-2 \citep{prsa16},
allowing for a very homogenous analysis of the full sample.

Ideally uniform space-based high-precision medium-cadence photometry and
homogenous high-resolution, high-stability spectroscopic ground-based
data for all the sample would suit best the purpose of the very precise SBC
calibration. Such light curves will become available for many of the systems here if the TESS mission is successfully launched.
We see this as a long-future next, natural step resulting in
additional improvements over internal consistency and precision of derived
physical parameters.

\subsection{Sample enlargement}
In order to increase the number of eclipsing binaries with suitable data for this programme, we have
selected a number of additional, suitable detached eclipsing
binary systems and have collected spectroscopic and photometric data
for them. They cover a wide range of spectral classes from B- to early
K-type and they are mostly within 300 pc from the Sun (low extinction
regime). Besides AL~Ari, for which we are already presenting derived
physical parameters and a paper describing full analysis will be
published soon \citep{kon17} our on-going analysis is at an advanced stage for about 20 more systems.

The sample will be expanded in the near future with systems having
more precise Gaia parallaxes within and beyond 300 pc horizon. More
systems will also join the sample from efforts of other research groups
investigating eclipsing binary stars as a large number of high quality
light curves from ground based surveys (e.g. Super-WASP, soon the LSST)
and space based surveys (e.g. Kepler-2, soon TESS) is registered for
both known and newly discovered systems. Those efforts will
surely result in enlarging significantly the sample to about 100 systems
covering B-, A-, F- and G-type stars. More systems will not only help
to reduce statistical errors of relations but also to determine
the intrinsic spread of the SBC relations.

\subsection{Parallaxes}
\label{sub:par}
Future more precise Gaia parallaxes are fundamental for any significant 
improvement to SBC relations presented here. We forecast expected 
precision of Gaia parallaxes for the sample as follows. We assumed 
conservatively that the precision of astrometry for bright stars (3 mag $< G < 12$ mag) will 
be 15 $\mu$as and that the photocenter movement of an eclipsing 
binary will be unequivocally detected and taken into account
when it is larger than 35 $\mu$as. 
The resulting expected mean relative precision of Gaia parallaxes 
will be 0.6\% for our sample. Systematic uncertainty introduced into the
prediction of angular diameters will likely be smaller 
but to be conclusive on this point we need to wait for a final Gaia
release quality evaluation. Figure~\ref{fig:gaiafut} presents the expected
angular diameters of stars in our sample after the final Gaia data release,
assuming the same radii and uncertainties as in Tab.~\ref{tab:fot}.
Inspection of this figure suggests that 
much improvement is expected, especially for blue stars. Angular diameters with a sub-percent
precision will be available for more than half of all components in our eclipsing
binary sample. We add that the sample will be augmented by very precise
dynamical parallaxes from interferometric orbits for a number of long
period eclipsing binary stars.

\subsection{Disentangling of component magnitudes}
\label{sub:dis}
It is interesting to estimate to what extent our extrapolation procedure introduce a bias. 
As was mentioned already in Sec.~\ref{sec:intrin},
flux ratios are calculated using precomputed intensities based on
ATLAS9 atmosphere models which assume a plane-parallel geometry and
local thermodynamic equilibrium (LTE). When components have similar
effective temperatures to within about 100 K their light ratio changes very 
little over the optical and NIR range of the spectrum and, regardless 
of the adopted model atmosphere, the extrapolation leads to negligible 
errors in comparison with observational photometric uncertainties. 
However the situation is somewhat different when the temperature 
difference between the components is much larger, say of the order of 1000~K.

For A-, F- and G-type stars with given atmospheric parameters ($T_{\rm
eff}, \log{g}$, [Fe/H]) and solar-like compositions their absolute
spectral energy distributions predicted by various atmospheric
models (plane parallel, spherical, LTE and non-LTE) in a range
of $B$ and $K$ bands have differences between them of up to 5\%,
but significantly smaller regarding relative fluxes (i.e. colors)
\citep[e.g.][]{bes98,mar07,edv08,ple11}. Comparison of model fluxes
with empirical fluxes in the aforementioned range of the spectrum gives
also very good agreements. As a result, we can expect, on average, a small systematic
uncertainty in the derived colors (reaching up to 0.02 mag) even in cases of
larger temperature difference between the components. Such an error would
be only a fraction of the typical uncertainty of an intrinsic color. This
uncertainty can be mitigated even further by using multi-band photometry and
carefully determined temperatures derived from disentangled spectra. For
hotter stars (O- and B-type) use of plane-parallel and LTE models may
lead to much larger systematic shifts \citep[e.g.][]{auf98,cug12},
however these issues will be addressed in a forthcoming paper.

\subsection{Photometry and transformations into standard system}
\label{sub:opt}
\subsubsection{Optical}
The precision of transformation
between the Tycho-2 and Johnson photometric systems is about 1\%
\citep{bes00} resulting in additional systematic uncertainty in our
SBC relations. To mitigate the problem one would use original Tycho-2
$B_T$ and $V_T$ magnitudes and to express the calibration in this
system. However we notice that for a few systems in our sample (EW~Ori,
VZ~Hya, VZ~Cep, LL~Aqr and EF~Aqr), Tycho-2 photometry transformed into
$B,V$ magnitudes give optical and NIR colors which are inconsistent with
each other and with temperatures of the stars. In those cases we used
other sources of $V$-band magnitudes. The source of discrepancy is unclear
to us, but we think that although the Tycho-2 photometry is multi-epoch in
particular cases mean $B_T$ and $V_T$ magnitudes are affected by the presence
of eclipses and/or other kind of systematics (e.g. transformation
errors). That strengthens the case for well-calibrated, precise and
uniform optical $B,V$ photometry in the standard Johnson system for stars in
the sample. In the optical, provided that a photometric system is close to the
standard one, it is expected that transformation from instrumental system
to the standard one would not produce systematic errors larger than 0.5\%.    

\subsubsection{NIR}
For the overwhelming majority of eclipsing binary
systems, well calibrated photometry NIR comes only from the single-epoch
2MASS survey. We transformed 2MASS magnitudes into the Johnson system
which may introduce systematics of up to 1\% (0.02 mag), because of
poor definition of the Johnson system in NIR. As an example of this fact
the transformation equation for $(V\!-\!K)$ color used by \cite{hol07}
has an offset of $-0.02$ mag with respect to the transformation equation
we used, of course a non-negligible value when we deal with sub-percent
precision. Preferentially the future SBC calibration should be expressed
in the 2MASS photometric system which is well calibrated \cite[e.g.][]{coh03}
and it is based on all-sky network of standard stars, or eventually
by using other NIR system which have similar bandpasses and precisely
determined transformation (e.g. SAAO).

Single-epoch photometry is prone to some accidental errors and
the statistical uncertainty of one measurement is relatively large. Because
of that it would be advisable to carry out new, high quality multi epoch
NIR photometry secured for stars in the sample. It would significantly
help in reducing statistical uncertainties and in removing any accidental
photometric errors. We already started a campaign to secure NIR photometry
for southern and equatorial stars from the sample with the plan to
derive precise out-of-eclipse magnitudes and later also to provide full
NIR light curves for some eclipsing binaries, especially those having
large effective temperature difference between components.

\subsection{Quantifying error contributions}
\label{sub:errors}
\begin{itemize}
\item{Radii: the mean precision of stellar
radii determination in our sample is 1.2\%. 
By using about 100 systems it would be
possible to pin-down the statistical error by a factor of 10, i.e to 0.1-0.2\%. 
Systematics will come mostly from the numerical tools for the analysis of eclipsing binary stars,
and it is expected to be of order of 0.1\%.}
\item{Parallaxes: taking into account the photocenter movements of the eclipsing 
binaries, the mean expected precision of Gaia parallaxes would be 0.6\%. 
The systematic error is expected to be significantly smaller.}
\item{Disentangling of magnitudes: up to 0.01 mag of systematics in derived colors
and magnitudes translates into a 0.3\% mean systematic uncertainty in predicting
angular diameters. However, by if we were to use full NIR light curves
and/or by using equal-temperature systems, then
this error could be almost eliminated because it would be
possible to determine the NIR magnitudes directly.}
\item{Photometric zero-points and transformations: in best cases of well-defined 
photometric systems (Section~\ref{sub:opt}) we expect 0.7\% systematics in colors and magnitudes.}
\item{Interstellar extinction: 
\newline a) Total extinction: the reddeing is low for almost all our systems. 
When we assume a standard Galactic exctinction curve with $R_V=3.1$, 
it introduces only a little additional uncertainty of about
0.03~mag in the $(V\!-\!K)_0$ color. Because the reddening line is
largely parallel to the SBC relation, this translates into only a 0.006~mag
statistical uncertainty (0.3~\%) in predicting the angular diameter. 
\newline b) Reddening law: for about 25\% of stars in within 1kpc 
from the sun \citep[e.g.][]{fit07,kre12}, we expect deviations from the universal law.
$R_V$ can vary significantly, but mostly lies between 2.7 and 3.7 \citep[e.g.][]{gon12}. 
When this is not accounted for, it shows as an additional intrinsic scatter in the SBC relation 
that amounts to about 0.02 mag in some individual cases.}
\end{itemize}

\begin{figure}
\mbox{\includegraphics[width=0.49\textwidth]{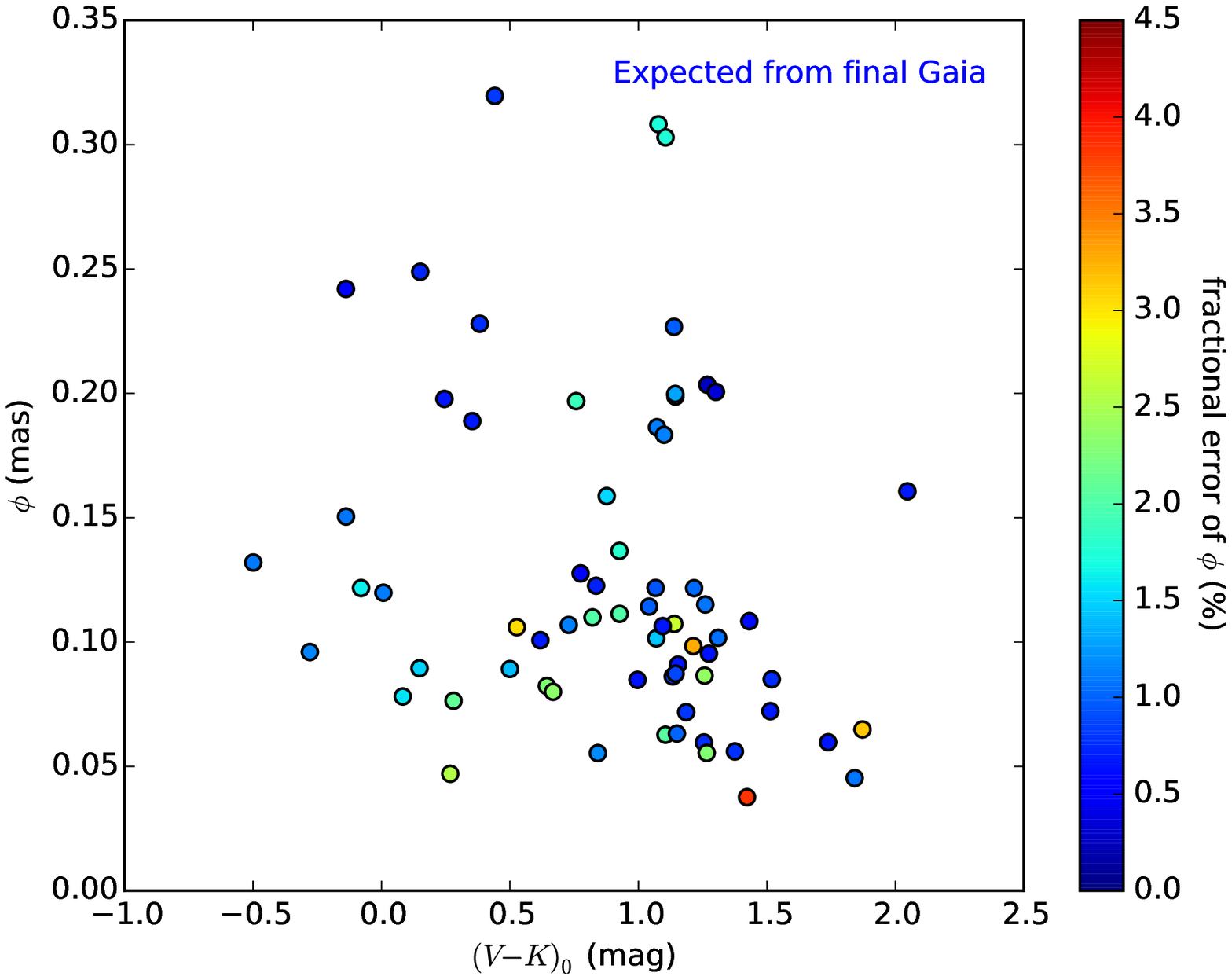}}
\caption{Predicted angular diameter uncertainties for stars in our
sample after the final Gaia release. Note the change in scale of the color bar with respect
to Fig.~\ref{fig:angular}. The error-bars would be in most cases smaller
than the size of the circles.}
\label{fig:gaiafut}
\end{figure}


The statistical uncertainty of the future SBC relation is expected to be well below 1\% provided
the number of suitable systems is sufficient (about 100 systems) and the internal
dispersion of a given relation is low. 
By combining all conservative estimates of errors from the above considerations
in quadrature we obtain an upper limit of 0.9\% on the systematic uncertainty. 
This error is dominated by the photometric uncertainties.

\section{Final Remarks}
New Gaia parallaxes combined with Hipparcos
and dynamical parallaxes allow us to derive for the first time the
SBC relations based fully on the eclipsing
binary stars. The precision of the derived relations for A-, F-, and
G-type stars is comparable to the precision of relations derived from
interferometric angular diameters, and both types of relations are
mutually consistent. The eclipsing binary method has no serious
limitations if it based on a well-selected sample of eclipsing
binaries, a self-consistent analysis method, and proper sanity
checks. To expand the SBC relation to O- and
B-type stars we propose to use the $B\!-\!K$ color, which allows reducing
interstellar extinction uncertainties. We also discussed all the steps necessary
to obtain precise and accurate SBC relations that allow for a
precision better than 1\% of the angular diameter predictions in the future.

\acknowledgments
The research leading to these results  has received
funding from the European Research Council (ERC) under the European
Union's Horizon 2020 research and innovation program (grant agreement
No 695099).

We are grateful for financial support from Polish National Science
Center grant MAESTRO 2012/06/A/ST9/00269. Support from the BASAL Centro
de Astrof{\'i}sica y Tecnolog{\'i}as Afines (CATA) PFB-06/2007, the
Millenium Institute of Astrophysics (MAS) of the Iniciativa Cientifica
Milenio del Ministerio de Economia, Fomento y Turismo de Chile, project
IC120009 and the IdP II 2015 0002 64 grant of the Polish Ministry of
Science and Higher Education is also acknowledged. We are also thanks
to the staffs in La Silla Observatory (ESO) and Las Campanas Observatory
(Carnegie) for their excellent support.

We also thank the anonymous referee for remarks and corrections to the text.

This research has made extensive use of the excellent astronomical
SIMBAD database and of the VizieR catalogue access tool, operated at CDS,
Strasbourg, France and made also use of NASA's Astrophysics Data System
Bibliographic Services (ADS).

This publication makes use of data products from the Two Micron All
Sky Survey, which is a joint project of the University of Massachusetts
and the Infrared Processing and Analysis Center/California Institute of
Technology, funded by the National Aeronautics and Space Administration
and the National Science Foundation.

We dedicate this work to Prof. Bohdan Paczy{\'n}ski who encouraged us many years ago 
to work on this subject.

{}    

\begin{appendix}
\section{Temperatures and reddening}
\subsubsection*{V570 Per}
The temperature of the system V570 Per was determined from a model
atmosphere analysis of disentangled spectra \citep{tom08b}. Although
formal errors on the temperatures quoted by the authors are very small (lower
than 0.5\%), the intrinsic colors of the components $b-y$, $B-V$, $V\!-\!J$
and $V\!-\!K$, point to much lower temperatures (by about 300 K), unless
there is significantly larger interstellar extinction to this object than
assumed by \cite{tom08b}: $E(B-V)=0.07$ mag instead of $0.023\pm 0.007$
mag. There are two ways of resolving the problem: (1) the temperatures are
indeed lower, or (2) the reddening is indeed higher. The first possibility
would force us to assume that some error was made by \cite{tom08b}
during their atmospheric analysis. This seems quite unlikely, however: (a)
their atmospheric analysis is standard, (b) the spectra are of good quality,
(c) higher temperatures correspond well with the components' spectral types
and masses. Thus the more probable explanation of disagreement is possibility
(2). However, it was reported that the interstellar potasium line KI (7699
\AA) is not detected in the spectra of the system, which would contradict
the higher reddening. Because we cannot solve this problem at the moment,
for the purpose of this work, we kept the temperatures from \cite{tom08b}
and assumed a reddening of $0.07\pm 0.03$ mag to V570~Per. This problem
clearly needs some future attention and more detailed investigation.    

\subsubsection*{WW Aur}
The temperatures of the components of WW~Aur were
previously determined by \cite{sma02} and \cite{sou05}. However,
$b\!-\!y$, $B\!-\!V$, $V\!-\!K$ colors suggest larger temperatures by
about $\sim 200$ K, what was already pointed out by \cite{sou05} in regard
of the $b\!-\!y$ color. \cite{wil09} used their direct distance estimate (DDE)
method and also found the temperatures of both components to be higher by a
very similar amount. In our model we employed those higher temperatures.

\subsubsection*{KX Cnc}
For the system KX~Cnc we determined the temperature
from Str\"omgren uvby photometry \citep[$b-y$=0.378;][]{ols83} and
Johnson's BVJK photometry. The temperature of the primary component
derived from the different colors is as follows: $T_{b-y}=5938$ K,
$T_{B-V}=5985$ K, $T_{V-J}=6131$ K, $T_{V-K}=6162$ K. The resulting
mean temperature is $T_1=6050$~K i.e. larger by 150 K (1.5$\sigma$) than
the original temperature $T_1$ derived by \cite{sow12}. The larger value is
in better agreement with the original HD spectral classification:
F8 \citep{can1919}. Using the calibration between effective temperature and
spectral type for normal main sequence stars \citep{pec13} we reclassify
the system as F9V+F9V.

\subsubsection*{RZ Cha}
 The case of RZ Cha is interesting. \cite{and75b}
combined their velocimetry with Str\"omgren photometry obtained by
\cite{jor75} to derive "mean" parameters of the components. The reason behind
it was their conclusion that the components of the system had very similar
physical appearance and thus also parameters. This "indistinguishability"
of components was retained by \cite{tor10} in their review. However, it
is clear from inspection of the light curves that the components have different
surface temperatures which was reported already by \cite{giu80}. The
difference is small, with the more massive and larger star being cooler by
$\sim 50$ K, but it has an effect on the predicted infrared light ratios.   

\subsubsection*{WZ Oph} 
 The system was quite recently analyzed by
\cite{cla08a}. They reported the temperature $T_1=6165 \pm 100$~K,
based on reddening E($B-V)=0.044$ mag, intrinsic Str\"omgren color of the
primary $(b-y)_0=0.329$ and a calibration by \cite{hol07}. They noted that
the temperature derived from atmospheric analysis of the disentangled primary's
spectrum is slightly higher, however they did not report how much
higher. From unreddened colors $b\!-\!y$, $B\!-\!V$, $V\!-\!K$ we derived
also a higher temperature of $T_1=6301$~K (1.4$\sigma$ difference).
Lower reddening of E($B-V)=0.030$ mag resulting from \cite{sch98} maps
leads to the temperature $T_1=6232$~K, a value which one would consider
"slightly" higher. These values of reddening and temperatures are assumed
in this work.     

\subsubsection*{UZ Dra}
Using $B,V,J,K$ photometry we redetermined
temperatures of both components because the original temperatures given
by \cite{slac89} were estimated only from the mean spectral type of the
system. Resulting temperatures are higher by about 200~K than those
reported by \cite{slac89} and correspond much better with the masses of both
components, which seem to be unevolved main-sequence stars.

\subsubsection*{VZ Cep}
There is an inconsistency between the temperatures
based on $B\!-\!V$, $b\!-\!y$ colors and $V\!-\!K$, $V\!-\!J$ with the NIR
colors resulting in temperatures higher by about 300~K. Different values
of reddening does not resolve the discrepant temperatures. A possible reason
is that 2MASS magnitudes are somehow affected, however they were taken
well outside of eclipses and all have an "A" flag. Higher temperatures would
be in agreement with relatively massive components of the system, and
furthermore the resulting photometric distance would be in perfect agreement
with Hipparcos and Gaia parallaxes. However, we have no clue at this
moment about the possible source of the discrepancy. We therefore retained in this paper 
the temperatures from the work by \cite{tor09}, which are based on 
Str\"omgren photometry.

\section{Radial velocities}

\subsubsection*{V570 Per}
\cite{tom08b} did not report radial velocity
semiamplitudes. We utilized data from their Table~2 to rederive the
orbital parameters. Our semi-major axis is larger by 1.5$\sigma$
than value by \cite{tom08b} which we attribute mostly to a different choice
of astrophysical constants, but our mass ratio $q$ is fully
consistent with their value.

\subsubsection*{HD 71636}
\cite{hen06} reported two sets of radial
velocity semiamplitudes in their Table~3 and Table~5 that contradict
each other. Thus we rederived the spectroscopic orbit from the data in their
Table~2. Our semiamplitudes are in perfect agreement with the
values presented in Table~3 and we accordingly adopted them here.

\subsubsection*{KX Cnc}
\cite{sow12} reported two sets of $K_{1,2}$
that contradict each other. Using the data from their Table~2 we determined
the spectroscopic orbit that is fully consistent with the solution given
in their Table~3.

\subsubsection*{V4089 Sgr}
Recently, \cite{ver15} presented light and
radial velocity curves solution of the system and they derived its absolute
dimensions. However, the semi-major axis $a$ reported in their Table~1 is
inconsistent with their radial velocity semiamplitudes $K_{1,2}$ and
masses. Our solution to the velocimetry kindly provided by M.~Veramendi
confirms their masses and  $K_{1,2}$, but not their $a$. Also our $K_2$
is slightly larger (by 1.2$\sigma$); this is probably caused by the fact
that we allowed for different systemic velocities for the
components. Finally, we recalculated errors on the fundamental parameters which are significantly
different from those reported in the Tables~1 and 2 by \cite{ver15}.

\subsection*{EF Aqr}
In paper by \cite{vos12} were presented fundamental
physical parameters of the system. However, they reported two different
sets of radial velocity semiamplitudes $K_{1,2}$ (their Tables 4 and
8). Using their velocimetry we redetermined spectroscopic orbits for this
system. Our $K_{1,2}$ are much closer to the values presented in Table 8,
but they are still somewhat different. Especially the epoch of spectroscopic
conjuction is different in our solution by 0.002 days suggesting some
period change in the system. We also recalculated fractional radii from
the sum of radii and $k$ given in their Table 6. The resulting radii and
errors are again somewhat different from those reported in Table 6. 
Here we refer only to parameters that we have recalculated.

\subsubsection*{V821 Cas}
 \cite{cak09} reported their radial velocity
measurements of the system. Their data in Table~1 are relatively noisy
compared to preset-day standards, nevertheless, we rederived the spectroscopic
orbits in order to verify the consistency of the orbital parameters and
quoted errors. The radial velocity semiamplitudes from our solution are
marginally consistent with their values and the overall agreement of the
orbit is satisfactory, also regarding the assumed errors.

\end{appendix}

\end{document}